\documentclass[aps,prb,twocolumn,groupedaddress,floats,showpacs]{revtex4}
\usepackage{amssymb,amsmath}
\usepackage{latexsym}
\usepackage{graphicx}
\usepackage{epsfig}

\begin{document} 

\title{Three--body Correlation Effects 
on the Spin Dynamics 
of Double--Exchange Ferromagnets}

\author{M. D. Kapetanakis, A. Manousaki,  and 
I. E.  Perakis}

\affiliation{Department of Physics, University of Crete, 
and 
Institute of Electronic Structure \& Laser, Foundation
for Research and Technology-Hellas, Heraklion, Crete, Greece} 

\date{\today}

\begin{abstract}
We present a  variational calculation 
of the spin wave 
excitation spectrum of double--exchange ferromagnets
in different dimensions. 
Our theory recovers 
the Random Phase 
approximation  and  1/S  expansion results 
as limiting cases and can be used to study the intermediate exchange 
coupling and electron concentration regime relevant to the manganites.  
In particular, we treat exactly the 
long range 
three--body correlations between 
a Fermi sea electron--hole pair and  a magnon excitation
and show that they strongly affect 
the spin dynamics in the parameter range 
relevant to experiments in 
the manganites. 
The manifestations of these correlations are many-fold. 
We demonstrate that they 
significantly change  the ferromagnetic phase boundary.
In addition to a  decrease in the 
magnon stiffness, we obtain an 
instability 
of the ferromagnetic state against spin wave excitations 
close to the 
Brillouin zone boundary.
Within 
a  range of intermediate 
concentrations,
we find a 
strong softening of the 
spin wave  dispersion as compared to 
the Heisenberg ferromagnet with the same stiffness, 
which changes into hardening 
for other concentrations. 
We discuss the relevance of these results 
to experiments in  colossal magnetoresistance  ferromagnets. 
\end{abstract}


\maketitle

\section{Introduction and problem setup}
\label{sec:intro}

The 
interactions between  itinerant carriers and 
local magnetic moments lead to new magnetic 
 properties in a wide variety  of  systems 
 that have been the subject of intense research 
lately. Examples where such 
magnetic exchange and Hund's rule interactions
play an imporant role 
range from ferromagnetic semiconductors such as EuO, EuS, 
chrome spinels, or pyrochlore \cite{nagaev} to  dilute III-Mn-V and 
II-VI magnetic 
semiconductors. \cite{III-V-1,III-V-2}
Of particular interest here are the manganese oxides (manganites)
R$_{1-x}$A$_{x}$MnO$_{3}$, where R=La, Pr, Nd, Sm, 
$\cdots$ and  A= Ca, Ba, Sr, Pb, $\cdots$,   
which display 
colossal magnetoresistance. \cite{tokura,dagotto}
The above  systems display ferromagnetic order
mediated by 
itinerant carriers.
Our main goal in this paper is to describe  the role 
of ubiquitous three--body correlations 
(beyond the mean field approximation)
on the spin dynamics 
of such ferromagnets. 

Given the wide variety of such  ferromagnetic systems, 
it is important to 
understand the properties of the minimal Hamiltonian that describes
the common properties and spin dynamics.  
In order to calculate  the 
effects of correlations 
beyond the mean field approximation, 
it is often necessary 
to neglect particularities of the individual systems,
such as  chemical structure and crystal environment.  
The most basic model that applies to all such materials 
is the Kondo lattice or double exchange Hamiltonian 
$H = K + H_{exch} + H_{super}+H_{U}$, where   
$K = \sum_{{\bf k} \sigma} 
\varepsilon_{{\bf k}} c^\dag_{{\bf k} \sigma} 
c_{{\bf k} \sigma}$
is the kinetic 
energy of the itinerant carriers.
In the manganites (R$_{1-x}$A$_{x}$MnO$_{3}$), $n=1-x$ 
itinerant electrons per Mn atom
occupy the band of 
Mn d--states with e$_{g}$ symmetry.
We simplify the calculation of the three--body correlations 
of interest here 
by considering a single 
tight--binding band 
of cubic symmetry
and neglect the bandstructure and the degeneracy of the e$_{g}$ states
(which may be lifted by external perturbations). 
The operator  $c^\dag_{{\bf k} \sigma}$ 
creates an electron with momentum ${\bf k}$, spin 
$\sigma$, and energy 
$\varepsilon_{{\bf k}}=
-2 t \sum_{i =1}^{d} \cos k_{i} a$, 
where $d=1,2,3$ is the system dimensionality and $a$ is the lattice constant.
From now on we take $a=1$ and measure the momenta in units of 
$\hbar/a$.

A common feature of all the systems of interest here is the strong 
magnetic exchange interaction, $H_{exch}$, 
between the itinerant carrier spins 
  and the local 
spin--S
magnetic moments ${\bf S}_i$, which are 
located at the $N^d$ lattice
sites ${\bf R}_i$.
In  the manganites, 
these S=3/2 spins are due to the 
three electrons in the tightly bound t$_{2g}$ 
orbitals. Introducing  the collective localized spin operator 
${\bf S}_{{\bf q}} = 1/\sqrt{N} 
\sum_{j} {\bf S}_{j} e^{- i {\bf q}{\bf R}_j }$
and the corresponding spin lowering 
operator $S^{-}_{{\bf q}}=S^{x}_{{\bf q}} - i S^y_{{\bf q}}$, 
we  express the local magnetic exchange interaction 
in momentum space:
\begin{eqnarray} 
\label{Hexch}
H_{exch} 
&=& -\frac{J}{2\sqrt{N}}\sum_{{\bf k} {\bf q} \sigma}
\, \sigma \, S_{{\bf q}}^z  \
c_{{\bf k}-{\bf q}\sigma}^\dag c_{{\bf k} \sigma} \nonumber \\
& &-\frac{J}{2\sqrt{N}}\sum_{{\bf k} {\bf q}}
 \left( S_{{\bf q}}^- \ c_{{\bf k}-{\bf q} \uparrow}^\dag 
 c_{{\bf k} \downarrow}+ h.c.\right), 
  \end{eqnarray}
where $\sigma = \pm 1$.
In the  manganites, $J>0$ describes the ferromagnetic Hund's rule 
coupling between the local and itinerant spins on each lattice 
site. 
$H_{super}$ is the 
weak antiferromagnetic 
direct superexchange interaction between the spins localized in neighboring 
sites, while $H_U$ describes the local Coulomb (Hubbard) 
repulsion among the itinerant electrons. 
The precise values of the parameters entering in the above Hamiltonian 
are hard to calculate for strongly coupled many--body 
systems  such as the manganites. 
Although the parameter estimates vary in the literature, 
typical values  
are $t \sim$ 0.2-0.5 eV and  
$J \sim$ 2eV, which corresponds to 
$4 \le J/t \le 10$. \cite{dagotto} On the other hand, 
the antiferromagnetic superexchange interaction 
is weak, $\sim 0.01 t$. 
The electron concentration, $n=(N_e/N)^d=1-x$ where $N_e$ is the 
number of electrons,
varies from 0 to 1. Ferromagnetism in the metallic state 
is observed within a concentration range 
$0.5 \le n \le 0.8$ in both 3D and quasi--2D (layered) systems. 
In this paper we neglect for simplicity the effects of $H_{super}$ 
and $H_U$ (to be studied elsewhere) in order to focus 
$H_{exch}$.

Given the large values of $J/t$ in most systems of interest,
a widely used approximation is the $J \rightarrow \infty$ 
limit
(double exchange ferromagnet). \cite{de}
In this strong coupling limit, 
the itinerant carrier 
is allowed to hop on a site only if  
its spin is parallel to the local spin on that site. 
The kinetic energy 
is then reduced when all itinerant and local 
 spins are  parallel, 
which favors a ferromagnetic ground state (double exchange 
mechanism).
We denote this fully polarized half--metallic state by $|F\rangle$
and note that it is an eigenstate of our Hamiltonian $H$. 
This state describes local spins with $S_z=S$ on all lattice sites 
and a Fermi sea of spin--$\uparrow$ itinerant electrons occupying 
all momentum states 
with $\varepsilon_{{\bf k}} \le E_F$, 
where $E_F$ is the Fermi energy.

Another commonly used approximation is to treat 
the local spins  as classical 
($S \rightarrow \infty$  limit). \cite{dagotto}
The ferromagnetism can then be
described by    an effective 
nearest neighbor Heisenberg model with ferromagnetic interaction. 
The quantum effects are often taken into account 
perturbatively in 1/S. 
This 1/S  expansion can be implemented systematically 
by using the Holstein--Primakoff 
bosonization method. \cite{golosov,chubukov,nagaev}
To $O(1/S)$, this method gives 
noninteracting Random Phase Approximation (RPA) magnons, 
whose dispersion 
is determined by the exchange 
interaction. \cite{furukawa,RPA}
In the strong coupling limit $J/t \rightarrow \infty$, this 
RPA dispersion coincides with that 
of the nearest neighbor Heisenberg ferromagnet. 
The $O(1/S^2)$ correction to the spin wave dispersion 
however deviates from this Heisenberg 
form. This correction  comes from the  scattering 
of the RPA magnon with the 
spin--$\uparrow$ electron Fermi sea 
when  treated 
to lowest order in the
electron--magnon interaction strength (Born approximation).
\cite{golosov,chubukov}

The role of nonperturbative carrier--magnon correlations 
(beyond $O(1/S^2)$) 
has been studied 
by exact diagonalization of small and 1D systems \cite{num-1,num-2} 
or by using  variational wavefunctions \cite{okabe,wurth}  
inspired from the Hubbard model
and the Gutzwiller wavefunction.
\cite{roth,rucken,ander,linden}
The   variational calculations of Refs.\onlinecite{okabe,wurth}  
treat the local correlations  expected 
to dominate in the strong coupling limit. \cite{rucken} 
The ferromagnetic (Nagaoka) state $| F \rangle$
was shown to become unstable with increasing electron concentration
due to  the softening
of either single particle spin excitations \cite{ander} 
or long wavelength spin wave excitations
(negative stiffness).\cite{okabe,wurth}
The spin wave dispersion 
deviated  
from the Heisenberg form  for very small
electron concentrations.\cite{wurth}  
In the  concentration range 
$0.5 \le n \le 0.8$ 
relevant to the manganites,
Wurth {\em et.al.} \cite{wurth} 
found small deviations 
from the Heisenberg form. 
A similar conclusion was reached based on the 1/S expansion.\cite{chubukov}
For $n \sim 0.7$, the $O(1/S^2)$  magnon dispersion 
showed a relative hardening 
at the zone boundary 
in the strong coupling limit. 
\cite{chubukov} Given 
the questions raised in the literature about the adequacy of the 
simple double exchange model for explaining the magnetic and transport 
properties of  the manganites \cite{dagotto}, it is important to 
treat  the Hamiltonian $H$ 
in a controlled way in the parameter regime relevant to the manganites.
Such a treatment  
would allow us to assess the accuracy of the commonly used 
approximations and understand the successes 
and limitations of the very basic model in explaining the experiments.

Due to the interplay 
between the spin and  charge degrees of freedom, 
a good understanding of the spin dynamics
is important for understanding the physics of 
colossal magnetoresistance and transport 
in the manganites.
Several experimental studies of the spin wave excitation 
spectrum have been reported in the literature. 
Heisenberg--like magnons 
were observed in the ferromagnetic regime 
for high electron concentrations (typically 
$n > 0.7$). \cite{sw-exp-1}
This observation is consistent 
with the spin wave spectrum obtained in the classical spin limit 
or by using the strong coupling RPA.
However, for lower electron concentrations $0.5 \le n \le 0.7$, 
unexpectedly strong deviations from the short range Heisenberg 
magnon dispersion were observed 
in several different manganites. 
\cite{soft-exp-1,soft-exp-2,soft-exp-3,soft-exp-4,endoh,soft-2D-1,soft-2D-2}
Most striking is the pronounced softening 
of the spin wave dispersion and short magnon lifetime close to the 
zone boundary, 
which indicate a new spin dynamics in the metallic ferromagnetic 
phase  
for intermediate 
electron concentrations 
$0.5 \le n \le 0.7$. 
The physical origin of 
this dynamics 
remains under debate.
It has been conjectured that the coupling to additional 
degrees of freedom
not included in the double exchange Hamiltonian $H$ 
is responsible for this new spin dynamics. 
Some of the  mechanisms that have been proposed involve
the orbital degrees of freedom, the spin--lattice
interaction, the local Hubbard interaction,
bandstructure effects, 
etc.   
\cite{endoh,khal,golosov,solovyev,mochena}. 

In this paper we 
study variationally 
the low energy spin excitations of the half--metallic 
fully polarized state $| F \rangle$. Our focus is on the role 
of correlations.
Our theory treats 
exactly up to three--body correlations between a magnon 
and a 
Fermi sea pair
by using the most general variational wavefunction that 
includes up to one Fermi sea pair excitations. 
As already noted in the context of the Hubbard model, \cite{rucken,linden}
the Gutzwiller 
wavefunction, which treats local correlations, \cite{wurth,okabe}
is a special case of such a wavefunction. 
Here we treat 
both local and long--range 
correlations on equal footing in momentum space
in order to interpolate 
between the weak and strong coupling limits
with the same formalism. 
We treat  nonperturbatively in a  
variational way   the 
multiple electron--magnon and hole--magnon 
scattering processes  that lead to 
vertex corrections 
of the carrier--magnon interaction. 
The above two  
scattering channels 
are {\em coupled}  
by  three--body correlations.
We show that this coupling is  important for the intermediate 
electron concentrations and exchange 
interactions relevant to the manganites, while for small (large) 
$n$ the electron--magnon (hole--magnon) 
scattering channel dominates. 
In the case of the 1D Hubbard model, a similar three--body 
treatment 
gave excellent agreement with the exact results.
\cite{igarashi}
Analogous   calculations 
were  performed to describe 
the electron--Fermi sea pair  local Hubbard 
interactions 
\cite{igarashi,rucken,linden} and 
the valence (or core) 
hole--Fermi sea pair interactions 
that lead to the Fermi Edge ( X--ray Edge) 
Singularity. \cite{perakis,ssr}

Our variational wavefunction 
offers several advantages.
While  local correlations 
\cite{wurth,okabe} 
dominate in the 
strong coupling limit, long range 
correlations  become important
as  $J/t$ decreases.\cite{rucken}
By  working in 
momentum space, 
we treat both long and short range 
correlations while 
addressing both the weak and strong coupling  limits
with the same formalism.  
We therefore expect that our results interpolate well 
for the intermediate values of 
$J/t$ relevant to the manganites.
\cite{rucken}
Our 
wavefunction satisfies momentum conservation
 automatically, 
 which reduces the number of independent variational parameters. 
Furthermore, our results become  exact 
in the two limits of $N_e=1$ and $N_e = N^d$.
We therefore expect  that they interpolate well 
for intermediate electron concentrations $0 < n < 1$.
Our variational equations contain the RPA ($O(1/S)$), 
ladder diagram approximation, 
and $O(1/S^2)$ results 
 as special cases.
Finally, our results converge 
 with increasing system size $N$ 
 and thus apply to the 
thermodynamic limit. 
The only restriction 
is that we neglect contributions from 
two or more  Fermi sea pair excitations.
Such multipair contributions are however suppressed 
for large $S$, while their contribution in the case of the 1D Hubbard model 
was shown to be  small. \cite{igarashi,ssr}

Here we address a number of  issues regarding 
the effects of up to three--body carrier--magnon correlations 
on the spin dynamics predicted by the simplest double exchange 
Hamiltonian.
First, by 
comparing to 
the 
1/S expansion,  RPA,  and 
ladder approximation
results, we show that vertex corrections and 
long range three--body  magnon--Fermi sea pair 
correlations, which couple the electron--magnon and 
hole magnon scattering channels,
play an important role
on the spin dynamics in the parameter 
regime relevant to the manganites. 
We find large deviations from the 
strong coupling 
double exchange spin wave dispersion, including  
a strong magnon softening at the zone boundary
in the intermediate electron concentration and exchange interaction regime.  
On the other hand, 
for small electron concentrations (relevant e.g. in III(Mn)V 
semiconductors \cite{III-V-1,III-V-2}), 
the electron-magnon multiple scattering processes dominate. 
However, even in this regime the deviations 
from the RPA and $O(1/S^2)$ magnon dispersions can be strong. 

Second, by
using an unbiased variational wavefunction,
we determine the change in  the ferromagnetic phase boundary  
due to the three--body 
correlations and carrier--magnon vertex corrections
(not included  to $O(1/S^2)$). 
The variational 
nature of our calculation allows us to 
rigorously conclude that the ferromagnetic 
state 
$|F\rangle$
is unstable when its energy 
exceeds that of the 
variational spin wave 
energy. In addition to the 
long wavelength softening 
and eventual instability,
which occurs in all dimensions, 
we find another instability 
for momenta close to the zone boundary
while the stiffness remains positive. 
This instability only occurs in 2D and 3D
for intermediate 
electron concentrations 
($0.4 \le n \le 0.7$ for the 2D 
three--body calculation).
This effect is exacerbated by the three--body 
correlations. 
One should contrast the above instability 
to the spin wave softening (but not instability) 
at the zone boundary that occurs
for  small $n < 0.3$.\cite{chubukov,golosov,wurth}

Third, we study the deviations 
from the Heisenberg spin wave dispersion
induced by the three--body correlations and vertex
corrections. 
This comparison  is important 
given the experimental observation of pronounced  deviations for 
$n \le 0.7$. 
\cite{soft-exp-1,soft-exp-2,soft-exp-3,soft-exp-4,endoh,soft-2D-1,soft-2D-2}
Deviations
from  Heisenberg behavior already occur to 
$O(1/S^2)$,  
or even to $O(1/S)$ for finite $J/t$, 
but in most cases correspond to magnon hardening.\cite{chubukov}
By comparing our results to the Heisenberg dispersion with the 
same stiffness, 
we show that, for  values of $J/t$
relevant to the manganites
and such that the 
ferromagnetic state is stable up to $n \sim 0.8$
or higher, the three--body correlations  in the 2D system 
give magnon hardening at the zone boundary for $n \le 0.4$ 
followed by strong magnon softening for $0.4 < n  \le 0.7$ 
and then small 
magnon hardening for $n > 0.7$. 
This behavior is similar to the experiment. 

The outline of this paper is  as follows. 
In Section
II we discuss the four approximations that we use to calculate 
the effects of the carrier--magnon correlations on the spin wave dispersion. 
In Section  II.A we discuss the variational wavefunction 
that treats the three--body correlations.  
The variational equations that determine the spin wave dispersion are  
presented in Appendix I. 
We also obtain the RPA magnon dispersion  variationally. 
In Section II.B we establish the 
connection between the above variational results and the 1/S expansion 
results. \cite{chubukov,golosov} We show, in particular, that the 
$O(1/S^2)$ magnon dispersion can be obtained from our variational equations 
by treating the carrier--magnon scattering to lowest order in the 
corresponding interaction strength (Born approximation). 
In Section II.C we discuss the two--body ladder approximation, 
obtained from our 
variational results by neglecting the coupling between
the  electron--magnon and hole--magnon scattering 
channels. The latter coupling is discussed further in Appendix II. 
In Section II.D we 
discuss the approximation of 
carrier--localized spin scattering 
and show that this variational  treatment 
improves on the RPA while 
making the numerical  calculation of three--body 
effects feasible in much larger 
systems. 
In Section III we present our numerical results for the 
spin wave dispersion, ferromagnetic phase diagram, 
and deviations from Heisenberg dispersion in the 1D, 2D, 
and 3D systems and compare between the different approximations. 
We end with the 
conclusions. 

\section{ Calculations } 
\label{wave} 

In this section we discuss 
the four approximations that we use to treat the effects 
of the carrier--magnon correlations. 
From now on we measure 
the energies $\omega_{{\bf Q}}$ of the spin wave states with respect 
to  that of
the fully polarized (half--metallic) ferromagnetic state $| F \rangle$,
whose stability and low energy spin excitations we wish to study.  
We note that $| F \rangle$ 
is an exact eigenstate of the Hamiltonian $H$ 
with maximum spin value and total spin z--component   $N(S + n/2)$. 
To study the stability of this
state 
in a 
controlled  way, 
it is important to use  approximations that 
allow us to draw definite  conclusions.
This is possible with the variational 
principle, which  allows us to conclude 
that a negative  excitation energy
$\omega_{{\bf Q}}$  
means instability of  
$| F \rangle$,
driven by the spin wave of momentum 
${\bf Q}$.  
Our variational states 
have the form 
$|{\bf Q} \rangle = 
M^{\dag}_{{\bf Q}}
| F \rangle$, 
where the operator 
$M^{\dag}_{{\bf Q}}$ 
conserves the total momentum, 
lowers the z--component of the total spin  by 1, 
and includes up to one
Fermi sea pair excitations. 
A spin wave has total
spin z--component of $N(S + n/2)-1 $, which corresponds to 
 one reversed  
spin as compared to $| F \rangle$. 
This spin reversal can be achieved 
either by lowering the localized spin z--component by 1
or by coherently promoting an electron from the  spin--$\uparrow$ 
band to the spin--$\downarrow$ band.
The spin reversal can be accompanied 
by  the scattering (shakeup)  of Fermi sea pairs. 
From now on we use the 
indices ${\bf \nu, \mu}, \cdots$ 
to denote single electron states inside the Fermi surface 
and  ${\bf \alpha, \beta}, \cdots$ 
to denote states 
outside the Fermi surface. 

\subsection{Three--body Correlations} 

In this section we discuss our three-body variational calculation 
of the spin wave dispersion.
First, however, we show that 
the well known RPA magnon dispersion \cite{furukawa,RPA}
can be obtained variationally 
for any value of $J/t$ 
by neglecting in $M^{\dag}_{{\bf Q}}$ 
all Fermi sea pair
excitations. The most general such operator has the form 
\begin{equation} 
\label{wav-RPA}
M^{\dag}_{{\bf Q}RPA}
= S^{-}_{{\bf Q}} | F \rangle 
+ \frac{1}{\sqrt{N}}\sum_{{\bf \nu}} 
X_{{\bf \nu}}^{{\bf Q} RPA} c^\dag_{ {\bf Q} + {\bf \nu} \downarrow}
c_{{\bf \nu} \uparrow}, 
\end{equation} 
where the variational equations for the $N_e$ amplitudes 
$X_{{\bf \nu}}^{RPA}$ 
are obtained in  Appendix I
(Eq.(\ref{RPAX})). 
In the strong coupling limit 
$J S \rightarrow \infty$,  
$X^{RPA} \rightarrow 1$ and $M^{\dag}_{{\bf Q}RPA}$
reduces to 
the total spin operator.
To lowest order in  $t/JS$ 
we obtain from 
Eqs. (\ref{RPAen}) and (\ref{RPAX}) 
that the  RPA dispersion then reduces to 
the  Heisenberg dispersion 
\begin{eqnarray} 
  \omega_{{\bf Q}}^{RPA}  = 
\frac{1}{2N} \frac{1}{S + n/2}
\sum_{{\bf \nu}<k_F} 
\left(\varepsilon_{{\bf v \nu} +{\bf Q}}
- \varepsilon_{{\bf \nu}} \right) + O(t/JS).  
\label{RPA-sc}
\end{eqnarray} 
The $O(1/S)$ magnon dispersion \cite{chubukov,furukawa} is obtained from 
the above strong coupling RPA result 
by replacing  the total spin prefactor $S + n/2$ 
by $S$. 

We now include in $M_{{\bf Q}}^\dag$ 
the most general contribution 
of the one Fermi sea pair states:  
\begin{eqnarray} 
\label{magnon}
&& M_{{\bf Q}}^\dag = S_{{\bf Q}}^{-}
      + \frac{1}{\sqrt{N}}
 \sum_{{\bf \nu}} 
  X_{{\bf \nu}}^{{\bf Q}} \
      c_{{\bf Q}+ {\bf  \nu} \downarrow}^{\dag}
      c_{{\bf \nu} \uparrow} 
+       \sum_{{\bf \alpha} {\bf \mu}} 
      c_{{\bf \alpha} \uparrow}^{\dag}
     c_{{\bf \mu} \uparrow} 
\times 
 \nonumber \\
      & &
\left[ 
\Psi^{{\bf Q}}_{{\bf \alpha} {\bf \mu}} \ 
S_{{\bf  Q} + {\bf \mu} - {\bf \alpha}}^{-}
\right. 
+ \left. \frac{1}{2 \sqrt{N}} \sum_{{\bf \nu}}
       \Phi^{{\bf Q}}_{\alpha \mu \nu} 
      c_{{\bf Q}+  {\bf \mu} - {\bf \alpha} + {\bf \nu} \downarrow}^{\dag} 
      c_{{\bf \nu} \uparrow} \right],  
  \end{eqnarray}
  where the amplitudes  $ X^{{\bf Q}}_{{\bf \nu}}
  ,\, \Psi^{{\bf Q}}_{{\bf \alpha \mu}}$ and  
 $\Phi^{{\bf Q}}_{{\bf \alpha \mu \nu}}$
  are all determined variationally; 
 we do not use the RPA results for 
$ X^{{\bf Q}}$ and 
 $\Phi^{{\bf Q}}$. 
As compared to previous calculations, we do not 
assume any particular 
form or momentum dependence for the above 
variational amplitudes. 
This allows us to treat in an unbiased way 
the long range correlations for  any 
value of $J/t$. 
The first two terms on the rhs of Eq.(\ref{magnon}) 
create  a magnon of momentum ${\bf Q}$. 
The last two terms describe
the scattering 
of a momentum 
${\bf Q}$ magnon to momentum 
${\bf Q}+ {\bf \nu} - {\bf \alpha}$
with the simultaneous scattering 
of  a Fermi sea electron 
from momentum ${\bf \mu}<k_F$ 
to momentum ${\bf \alpha} > k_F$. 

Our wavefunction Eq. (\ref{magnon}) becomes exact 
in the two limits of $N_e=1$ and $N_e=N^d$. 
To see this, we note that, for $N_e=1$,  the 
Fermi sea consists of a single electron. As a result, 
multipair excitations
do not contribute,  
while $\Phi^{{\bf Q}}=0$.
In the half--filling limit  $N_e=N^d$, all lattice sites are occupied 
by one spin--$\uparrow$ electron
and the Fermi sea occupies all momentum states up to the zone boundary. 
As a result, the RPA wavefunction Eq. (\ref{wav-RPA}) becomes exact. 
Eq. (\ref{magnon}) also gives the  exact wavefunction 
in the atomic limit $t=0$, $\varepsilon_{k}=0$, where 
the variational amplitudes do not depend on the 
electron momenta. 
To see this, we note that, due to the Pauli principle, 
$\Phi_{{\bf \alpha \mu \nu}}^{{\bf Q}}$ must be antisymmetric with respect 
to the exchange of 
the Fermi sea electron momenta $\nu$ and $\mu$. 
In the atomic limit, $\Phi^{{\bf Q}}$ must therefore vanish since it 
is independent of the momenta.  
For the same reason,  all multipair amplitudes vanish as well and  
Eq.(\ref{magnon}) 
gives the exact result.

The variational equation for $\Psi^{{\bf Q}}$ 
is derived in Appendix I (Eqs.(\ref{Psi-var}) and (\ref{Psi})).
The magnon energy 
is obtained 
from Eq. (\ref{omega-var}) after 
substituting Eq. (\ref{X-var}): 
\begin{eqnarray} 
\label{mag-en} 
\omega_{{\bf Q}} = \frac{J}{2N} 
\sum_{{\bf \nu}} 
\frac{\varepsilon_{{\bf \nu} + {\bf Q}}- 
\varepsilon_{{\bf \nu}}- \omega_{{\bf Q}}}{JS + 
\varepsilon_{{\bf \nu} + {\bf Q}}- 
\varepsilon_{{\bf \nu}}- \omega_{{\bf Q}}}
+ \sum_{{\bf \nu}} \Gamma^e_{{\bf \nu}}.  
\end{eqnarray} 
where we introduced the electron vertex function 
\begin{equation} 
\Gamma^{e}_{{\bf \nu}} = 
\frac{J}{2N} \frac{\varepsilon_{{\bf \nu} + {\bf Q}}- 
\varepsilon_{{\bf \nu}}- \omega_{{\bf Q}}}{JS + 
\varepsilon_{{\bf \nu} + {\bf Q}}- 
\varepsilon_{{\bf \nu}}- \omega_{{\bf Q}}}
 \sum_{{\bf \alpha}'} 
\Psi^{{\bf Q}}_{{\bf \alpha}' {\bf \nu}}. 
\label{Ge}  
\end{equation} 
The first term in Eq.(\ref{mag-en}) gives the RPA contribution 
to the magnon energy. 
The second term is the 
carrier--magnon self energy contribution, 
determined by the electron  vertex function $\Gamma^e$. 
The latter satisfies Eq.(\ref{Ge-eq}), 
which describes the multiple electron--magnon  scattering 
contribution (ladder diagrams, two--body correlations) 
as well as the coupling to the hole vertex function, 
Eq. (\ref{Gh}), due to the three--body correlations. 
Below we discuss  three 
contributions to the full $\Gamma^e$: 
$O(1/S^2)$ (Born scattering approximation), 
ladder diagram (two--body carrier--magnon correlations), 
and the contribution due to carrier scattering 
with the localized spins.

\subsection{1/S Expansion} 
\label{sec-1/S}

In this section we make the connection with the Holstein--Primakoff 
bosonization treatment of the quantum effects. \cite{golosov,chubukov}
In particular, we show that the $O(1/S^2)$ magnon dispersion
\cite{golosov,chubukov}
can be obtained by solving 
Eq.(\ref{Ge-eq}) 
perturbatively,  to lowest order in the carrier--magnon interaction and 1/S.
 
First, we recall that classical spin behavior is obtained in 
the limit 
$S \rightarrow \infty$ with $J S$=finite. 
By expanding Eqs (\ref{Psi}), (\ref{Ge}) and (\ref{Gh})  
in powers 
of the small parameter 1/S ($J S$= finite)  we see that 
$\Psi^{{\bf Q}}= O(1/S)$, 
$\Gamma^e=O(1/S^2)$, 
and $\Gamma^h=O(1/S^2)$. 
In particular, we obtain from Eq.(\ref{Ge-eq}) 
to lowest order in 1/S
\begin{eqnarray} 
\label{Ge-S2} 
\Gamma^{e}_{{\bf \nu}} 
= - \frac{J^2}{4N^2} \left( \frac{\varepsilon_{{\bf \nu} + {\bf Q}}- 
\varepsilon_{{\bf \nu}}}{JS + 
\varepsilon_{{\bf \nu} + {\bf Q}}- 
\varepsilon_{{\bf \nu}}} \right)^2 
\sum_{{\bf \alpha}}  
\frac{1}{
\varepsilon_{{\bf \alpha}}
- \varepsilon_{{\bf \nu}}}
, \ \ 
\end{eqnarray}  
and from 
Eq.(\ref{mag-en}) 
\begin{eqnarray} 
\label{mag-en-1/S} 
\omega_{{\bf Q}} 
&= &\frac{J}{2N} 
\sum_{{\bf \nu}} 
\frac{\varepsilon_{{\bf \nu} + {\bf Q}}- 
\varepsilon_{{\bf \nu}}-\omega_{{\bf Q}}}{JS + 
\varepsilon_{{\bf \nu} + {\bf Q}}- 
\varepsilon_{{\bf \nu}}- \omega_{{\bf Q}}}
\nonumber  \\
& & - \frac{J^2}{4 N^2}  \sum_{{\bf \alpha} {\bf \nu}} 
\frac{\left( \varepsilon_{{\bf \nu} + {\bf Q}}- 
\varepsilon_{{\bf \nu}} \right)^2}{\left(JS + 
\varepsilon_{{\bf \nu} + {\bf Q}}- 
\varepsilon_{{\bf \nu}}\right)^2
\left( \varepsilon_{{\bf \alpha}} - \varepsilon_{{\bf \nu}}
\right)}. \ \ 
\end{eqnarray} 
The last term in the above equation
comes from the lowest order 
magnon--electron 
scattering contribution.
The $O(1/S)$ 
spin wave dispersion \cite{furukawa}  is obtained 
from the first term 
by neglecting  $\omega_{{\bf Q}}=O(1/S)$ in the denominator.
The spin wave energy
to $O(1/S^2)$
is obtained by expanding the first term to this order. 
We recover  the strong coupling $O(1/S^2)$ results 
of Refs \onlinecite{golosov,chubukov}, 
obtained by using the bosonization technique, 
by further expanding Eq.(\ref{mag-en-1/S})  
in the limit 
 $J S \rightarrow \infty$.

The $O(1/S^2)$ magnon dispersion is not variational.
Thus we cannot definetely conclude instability of the ferromagnetic 
state if we find a negative magnon energy to $O(1/S^2)$. 
On the other hand, the three--body calculation 
outlined in the previous section treats the 
magnon--Fermi sea pair interaction variationally rather than perturbatively
(as in the 1/S expansion) 
while  recovering the 
$O(1/S^2)$ results
as a special case. 
The n--pair 
contributions to Eq.(\ref{magnon})  have amplitudes 
of order $O(1/S^n)$. Therefore,
 the shake--up of multipair excitations is suppressed 
for large $S$.
Our three--body calculation thus puts the 
$O(1/S^2)$ results on a more quantitative (variational) basis 
by treating fully rather than perturbatively all contributions 
of the one Fermi sea pair states.

adaa

\subsection{Two--body Ladder  Approximation} 
\label{ladder}

To go beyond the Born approximation ($O(1/S^2)$),  
we first consider the two--body correlation 
contributions to the Fermi sea pair amplitude $\Psi^{{\bf Q}}$
while still neglecting the three--body correlations.
This is equivalent  to treating the ladder diagrams
that describe the 
multiple electron--magnon and hole--magnon 
scattering, while neglecting the 
coupling between 
these two scattering channels. 
Noting 
that the magnon dispersion is determined by 
$\Gamma^e$ only (Eq.(\ref{mag-en})), 
the ladder approximation dispersion 
is  obtained from 
Eq. (\ref{Ge-eq}) with  $\Gamma^h=0$
and Eq.(\ref{mag-en}):
\begin{eqnarray} 
\label{mag-en-ladd} 
\omega_{{\bf Q}} = \frac{J}{2N} 
\sum_{{\bf \nu}} 
\frac{\varepsilon_{{\bf \nu} + {\bf Q}}- 
\varepsilon_{{\bf \nu}}- \omega_{{\bf Q}}}{JS + 
\varepsilon_{{\bf \nu} + {\bf Q}}- 
\varepsilon_{{\bf \nu}}- \omega_{{\bf Q}}} \nonumber \\ 
+ \frac{J^2}{4N^2} \sum_{{\bf \nu}} 
\frac{ 
 \left( \frac{\varepsilon_{{\bf \nu} + {\bf Q}}- 
\varepsilon_{{\bf \nu}}- \omega_{{\bf Q}}}{JS + 
\varepsilon_{{\bf \nu} + {\bf Q}}- 
\varepsilon_{{\bf \nu}}- \omega_{{\bf Q}}} \right)^2 
\sum_{{\bf \alpha}'}  
1/\Delta^{{\bf Q}}_{{\bf \alpha}' {\bf \nu}}}{ 
 1 - 
\frac{J}{2N} \frac{\varepsilon_{{\bf \nu} + {\bf Q}}- 
\varepsilon_{{\bf \nu}}- \omega_{{\bf Q}}}{JS + 
\varepsilon_{{\bf \nu} + {\bf Q}}- 
\varepsilon_{{\bf \nu}}- \omega_{{\bf Q}}}
 \sum_{{\bf \alpha}'} 
1/\Delta^{{\bf Q}}_{{\bf \alpha}' {\bf \nu}}},
\end{eqnarray} 
where 
$\Delta^{{\bf Q}}$ 
is given by Eq.(\ref{D}). We note that, similar to the 1/S expansion, 
the above ladder approximation result is not variational. 
A similar approximation 
was used  in the context 
of the Fermi Edge Singularity. \cite{perakis,ssr}
There it was shown that at least 
three--body correlations 
are necessary in order to 
describe the  unbinding of the 
discrete exciton bound state.  \cite{perakis,ssr}
In the case of the Hubbard model,  
the ladder approximation was shown to overestimate the 
electron self energy. \cite{igarashi}   

The difference between the spin wave dispersions Eq.(\ref{mag-en-ladd}) 
and the full three--body calculation comes from the 
three--body correlations. By comparing the corresponding 
curves in the next section we can therefore 
judge the role of these correlations on the spin dynamics.    

\subsection{Carrier--localized spin scattering ($\Phi=0,\Psi \ne 0$)} 

To  describe the three--body correlations, the coupled equations 
for $\Gamma^e$ and $\Gamma^h$
must be solved. Although this is possible 
in 1D and 2D for fairly large systems, 
in 3D the numerical solution of the 
full variational equations  
is challenging, due to the dependence of 
$\Gamma^h_{{\bf \alpha} {\bf \nu}}$ 
on six momentum components. 
On the other hand,  $\Gamma^{e}_{{\bf \nu}}$ 
depends on one momentum only.
The dependence of $\Gamma^h_{{\bf \alpha} {\bf \nu}}$ 
 on the momentum ${\bf \nu}$ can be eliminated by considering 
a simpler variational wavefunction, 
obtained from Eq.(\ref{magnon}) 
by setting  $\Phi=0$.
This corresponds to treating fully the scattering 
of the electron with the localized spins while neglecting the 
electronic contribution to the scattered magnon.
This approximation becomes exact in the two 
limits $N_e=1$ and $N_e=N^d$ and 
also recovers the $O(1/S^2)$ and RPA  results. 
Its main advantage is that it improves the RPA 
by allowing  us to 
treat variationally three-body 
carrier--localized spin  correlations in a large system. 
The corresponding spin wave dispersion is obtained by solving the 
coupled  Eqs.(\ref{Ge-eq}) and (\ref{Gh-eq}) and then substituting 
$\Gamma^e$ in Eq.(\ref{mag-en}). 
\begin{figure}
\centerline{
\hbox{\psfig{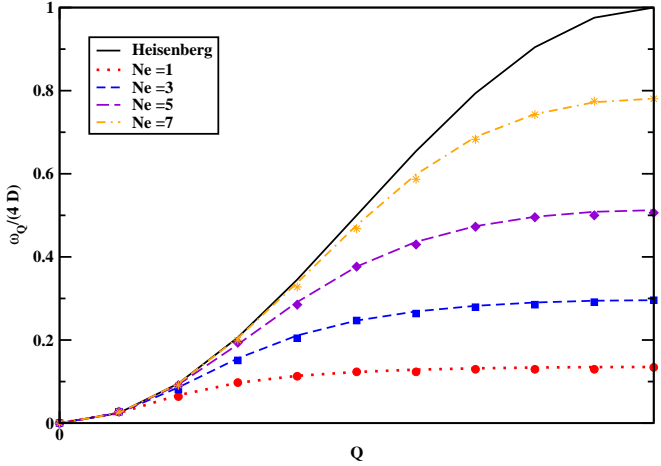}}
}
\caption{ Spin-wave dispersion  at the $J \rightarrow \infty$ limit in a 1D system with N =20, S =1/2
 and for number of electrons Ne =1, 3, 5, 7.
The points in the figure show exact results.  
}
\end{figure}

\section{Numerical results} 

In this section we present the results 
of our numerical calculations. 
First of all, we compare our numerical results with results made
by exact diagonalization of small 1D system\cite{num-1,num-2}. Fig 1
shows the comparison where we see that our results are in very good agreement
with the exact calculations.
To draw conclusions on the role of the correlations, 
we compare the 
different approximations discussed in the previous section 
for a d--dimensional lattice
with $N^d$ sites.
We perform our calculations for d=1, 2, and 3.
The dimensionality  
of the system affects the 
quantum fluctuations and  correlation effects.
Quantum fluctuations are expected to be most pronounced in the 1D system, 
where we show that the 1/S expansion can lead to spurious features.
The calculation of the 1D magnon dispersion 
could also  be relevant to quasi--1D materials with 
chain structures. 
Our 2D magnon dispersion is 
relevant to the  quasi--2D 
layered manganites, \cite{magnon-2D,2D-mang}
where a pronounced spin wave softening
and deviations from the Heisenberg dispersion 
similar to the 3D system
\cite{soft-exp-1,soft-exp-2,soft-exp-3,soft-exp-4,endoh}
were observed
experimentally. \cite{soft-2D-1,soft-2D-2} 
The similarity of the spin dynamics 
in the 3D and 2D systems 
indicates that the relevant physical mechanisms 
are  generic 
and do not 
depend crucially on the particularities of the individual 
systems.  
In 2D, the full three--body variational calculation can be performed 
in fairly large systems ($N \sim 20-30$), 
while in 3D it could only be performed for $N \sim 10$. 
Therefore, the 2D system also offers computational 
advantages.
On the other hand, the rest of the approximations discussed 
here can be performed in very large systems (up to $N \sim 200$),  
until full convergence with increasing $N$ 
is reached.

We start with the dependence of the spin wave excitation spectrum on the 
electron concentration $n$. 
Figs. 2 and 3 show the 
magnon dispersion in the 1D and 2D systems respectively 
for a fixed exchange intreraction,  $J/t=10$,
and four different values of $n$. 
The 2D dispersion (Fig. 3) was calculated along the Brillouin zone direction 
$(0,0)\rightarrow(\pi,0)$ ($\Gamma-X$), 
where the discrepancies between the different approximations are maximized. 
\begin{figure}
\centerline{
\hbox{\psfig{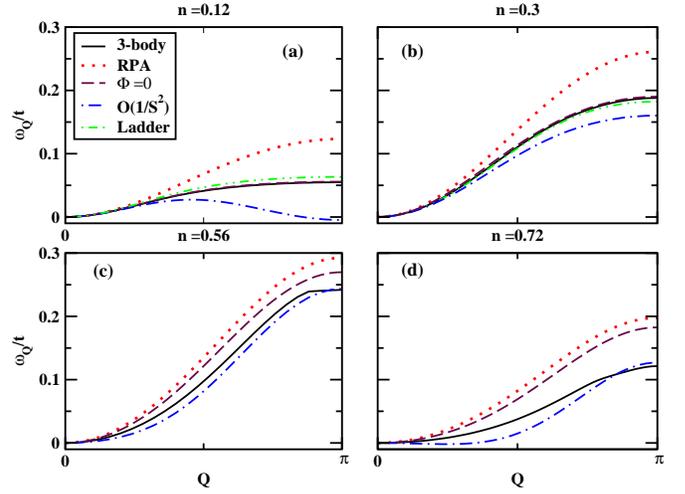}}
}
\caption{ Spin-wave dispersion  in the 1D system: 
comparison of the full three--body variational calculation (solid curve) 
to the 
different approximations discussed in the text.
$J/t=10$. 
}
\end{figure}

For very small 
electron concentrations 
($n=0.12$ in Figs. 2 and 3), 
the carrier--magnon scattering  
tends to soften the spin wave dispersion 
close to the zone boundary, 
consistent with previous results. \cite{wurth,chubukov} 
This can be seen by comparing in Figs. 2(a) and  3(a) 
the RPA dispersion with the different calculations  of the 
carrier--magnon scattering effects.
It is important to note that, despite the 
softening, 
the spin wave dispersion does not 
become negative (unstable) close to the zone boundary 
for very low concentrations. 
This softening may be interpreted as 
a remnant in the thermodynamic limit
of the failure of the RPA 
for $N_e=1$,
where Eq. (\ref{magnon}) gives the exact solution.
Indeed, for $N_e=1$,
the magnon energy is of
$O(1/N^2)$, while  the RPA gives   $O(1/N)$ energies.
Furthermore, for low
concentrations,  the 
$\Phi \ne 0, \Psi\ne 0$ and $\Phi=0, \Psi\ne 0$
variational calculations give results  similar to 
the  ladder approximation
(the corresponding curves almost overlap in Figs. 2(a) and 3(a)).
This indicates that the three--body correlations are weak 
for very low concentrations. 
This result can be understood  by noting that the 
last term in Eq. (\ref{Psi}) (and Eq. (\ref{D})),  
which describes the hole--magnon multiple scattering contribution, 
is suppressed for small $n$.  
Indeed, with decreasing  $n$ and Fermi energy $E_F$, the phase space 
available for the hole 
to scatter decreases relative to the phase space available 
for electron scattering. As a result, the electron--magnon 
scattering channel (electron ladder diagrams) 
dominates.
On the other hand, 
the difference between the above dispersions and the
RPA is strong, 
while the differences from the  
$O(1/S^2)$ (Born scattering) result
are  noticable even for very  small $n$ (Figs. 2(a) and 3(a)). 
The latter differences  come from 
the multiple electron--magnon 
and hole--magnon scattering processes. 
In 1D, the 
$O(1/S^2)$ result fails qualitatively 
for very low ($n=0.12$ in Fig. 2) 
and very high ($n \ge 0.8$) electron concentrations. 
For such concentrations, 
the $O(1/S^2)$ dispersion becomes negative (unstable) 
at the zone bondary. This zone boundary instability
persists  even in the strong coupling limit $J \rightarrow \infty$
but  
is absent in all our variational results. 
\begin{figure}
\centerline{
\hbox{\psfig{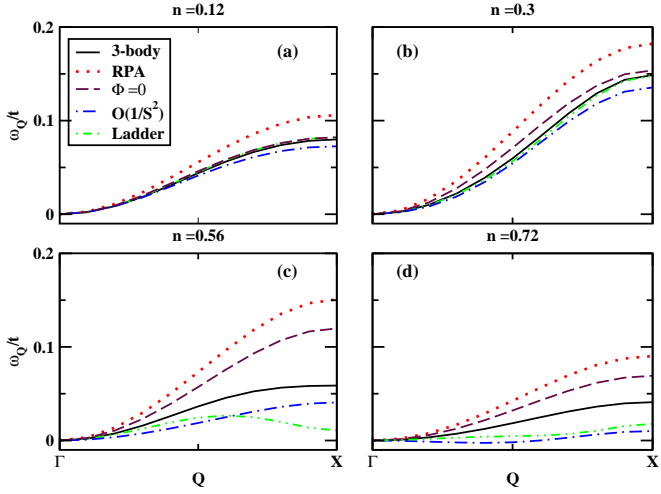}}
}
\caption{ Spin-wave dispersion in the 2D system along the direction 
$\Gamma-X$ for the same parameters as in Fig. 2: 
comparison of the full three--body variational calculation (solid curve) 
to the 
different approximations discussed in the text.
} 
\end{figure}

With increasing electron concentration, the 
spin wave energies and stiffness initially increase
(compare the $n=0.12$ and 
$n=0.3$ dispersions in Figs. 2 and 3). 
 Fig. 4 shows $D(n)$, obtained by fitting 
the quadratic behavior $D(n) Q^2$
to the long wavelength numerical dispersions, 
for finite exchange interaction $J/t=10$. 
As can be seen in Fig. 4, the RPA predicts  an initial increase of 
the spin wave stiffness
with $n$ followed by a decrease. However,    
the carrier--magnon scattering reduces the 
spin wave stiffness and 
changes the above  concentration dependence significantly, especially in  
2D and 3D (see Fig. 4). 
The difference from the RPA behavior is particularly striking for the 
$O(1/S^2)$ contribution to $D(n)$. The latter is significantly suppressed 
as compared to the rest of the approximations of carrier--magnon scattering. 
This particularly strong softening indicates that the $O(1/S^2)$ result 
significantly underestimates 
$T_c$ and the stability of the fully polarized ferromagnetic state. 
We note that, due to the variational nature of the full three--body 
calculation, the exact stiffness will be smaller (softer) 
than the  corresponding results of Fig.4 (solid curve). 

\begin{figure}
\centerline{
\hbox{\psfig{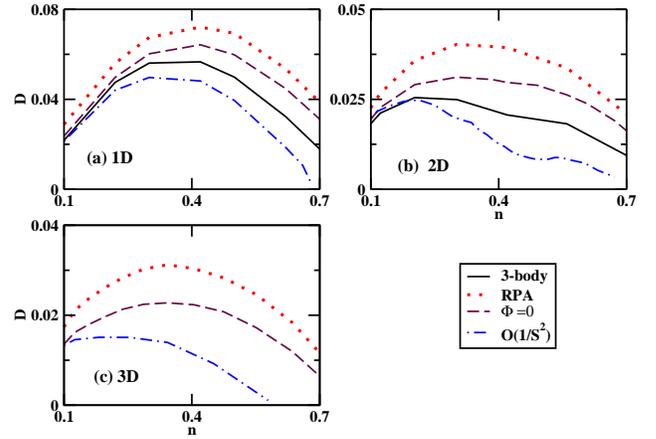}} 
}
\caption{
Spin wave stiffness for $J=10t$ as function of electron concentration
and system dimensionality. }
\end{figure}  

As the electron concentration increases, 
we see from Figs. 2 and 3 that the 
different approximations 
start to 
deviate substantially from each other. This is clear 
for the 2D system (Fig. 3),   
while in 1D the differences develop for  higher electron concentrations.  
Compared to the full three--body 
variational calculation ($\Phi \ne 0, \Psi \ne 0$), 
the ladder and $O(1/S^2)$ (non-variational) 
approximations give softer spin wave energies,
while the variational $\Phi=0, \Psi \ne 0$ and  RPA
($\Phi = \Psi = 0$) 
wavefunctions  
give higher spin wave energies.
The large differences between the above dispersions point out 
the importat role  of 
carrier--magnon correlations
for such electron concentrations. 
The difference between the full three--body calculation 
(or the $\Phi=0,\Psi \ne 0$ calculation) 
and the 
ladder approximation 
in Figs. 3(c) and 3(d) shows that 
the three--body correlations are signifcant,   
while the differences from the
$O(1/S^2)$ result show the importance  of the multiple carrier--magnon 
scattering processes (vertex corrections). 
One can conclude from the above comparisons that the
Fermi sea hole--magnon scattering cannot be neglected for such concentrations.
Furthermore, as can be seen in Fig. 3, 
the different aproximations 
bound the full three--body result. This is particularly useful for 
the  3D system, where the full three--body variational calculation 
could only be performed for 
relatively small  $N^d$ lattices with $N \sim 10$
(due to the  large number of variational parameters 
$\Psi_{{\bf \alpha} {\bf \nu}}^{{\bf Q}}$). 
On the other hand, 
Fig. 5 shows the spin wave dispersions for a
rather large  ($50^3)$ 
3D lattice, 
obtained by using the RPA, $O(1/S^2)$, and $\Phi=0,\Psi \ne 0$ 
approximations.  
By comparing the spin wave dispersions 
in Figs. 5 and 3, obtained 
for the same parameters, 
we see that the 
trends as function of $n$ are qualitatively similar 
in the 2D and 3D systems.

The differences between the  approximations 
studied here (which are due to the correlations) 
are very pronounced 
in the electron concentration range $0.5 \le n \le 0.8$
for intermediate exchange interaction values, i.e. in the parameter 
regime relevant  to the 
manganites.  
This can be seen more clearly in Fig. 6,
which shows the 2D spin wave  dispersions 
obtained with the different approximations 
along the main directions in the Brillouin zone 
for the typical parameters $n=0.7, J=8t$.
Fig. 6 compares the dispersions 
along the directions $(0,0) \rightarrow (\pi,0)$ ($\Gamma-X$), 
$(\pi,0) \rightarrow (\pi, \pi)$ ($X-M$), and along the 
diagonal $(0,0) \rightarrow (\pi, \pi)$ ($\Gamma-M$).  
The discrepancies between the different 
approximations are very large along $\Gamma-X$ 
but much smaller along the other directions. 
For example, 
the RPA fails completely 
along the direction $\Gamma-X$, where   
the full three--body calculation
shows a striking spin wave softening that is most 
pronounced close to the $X$ point.
Such a strong effect, much stronger 
than the softening  at small $n$,  
only occurs for 
intermediate electron concentrations 
$0.4 < n \le 0.7$ and is underestimated by the 
$\Phi=0, \Psi \ne 0$ variational calculation. 
On the other hand, 
the $O(1/S^2)$ dispersion for the parameters 
of Fig. 5 shows 
instability at long wavelengths (negative stiffness)
rather than softening at the zone boundary.

\begin{figure}
\centerline{
\hbox{\psfig{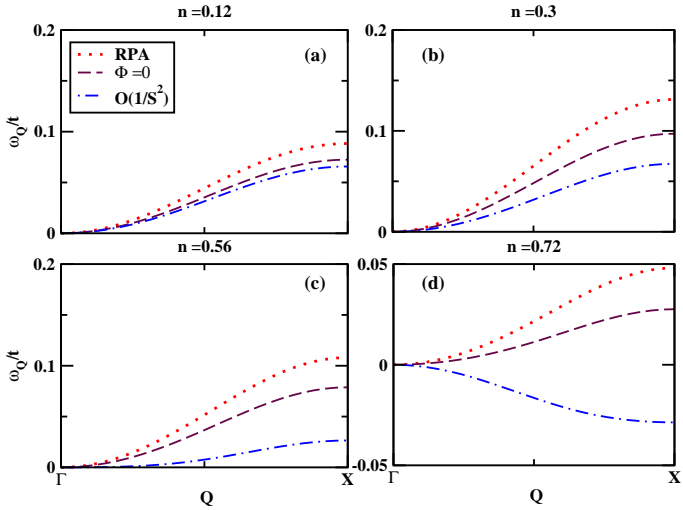}}
}
\caption{
Spin wave dispersions in 3D 
for $J/t=10$ and electron concentrations
similar to Figs. 2 and 3.  
}  
\end{figure}

To see the origin of the above spin wave softening,  
we show in 
Fig. 7 the spin wave dispersion for a slightly smaller 
$J/t$ than in Fig 6.
The spin wave energy now becomes negative 
in the vicinity of the $X$ point,  
while the magnon stiffness remains positive. 
This variational result 
allows us to conclude instability of the fully polarized ferromagnetic 
state due to the $X$ point magnons.
The strong zone boundary softening is a precursor to this instability. 
Fig. 7 also compares the full three--body and RPA calculations 
for two different values of $N$ with fixed $n$. 
For $n=0.7$,  our results have 
converged reasonably well even 
for $N \sim 10$ 
and thus reflect the behavior in the thermodynamic limit. 

The above zone boundary instability 
occurs  in 2D and 3D for intermediate 
electron concentrations ($0.4 \le n \le 0.7$ for the 2D 
three--body calculation), 
where a strong magnon softening and short magnon 
lifetimes were observed in the manganites.
\cite{soft-exp-1,soft-exp-2,soft-exp-3,soft-exp-4,endoh,soft-2D-1,soft-2D-2}
Although the other approximations 
can also give an instability, 
this occurs within a more limited range of $n$ and $J$ 
than for the 
full three--body calculation (discussed further below).   
The magnitude
and concentration dependence  of the softening 
also depends on the local Hubbard repulsion ($H_U$) and 
direct superexchange ($H_{super}$)
interactions 
and the bandstructure (to be considered elsewhere).
We note that spin wave softening at the zone boundary of electronic 
origin was obtained before 
within the one--orbital Hamiltonian for finite values of $J/t$
by including these additional  effects. \cite{golosov,mochena} 
The main difference here is 
that our calculation is variational (and thus allows us to draw 
definite conclusions by guaranteeing  
that the exact magnon energies are even softer than the calculated values)
and our effect was obtained by using the 
simplest possible Hamiltonian ($H_U=H_{super}=0$).

\begin{figure}
\centerline{
\hbox{\psfig{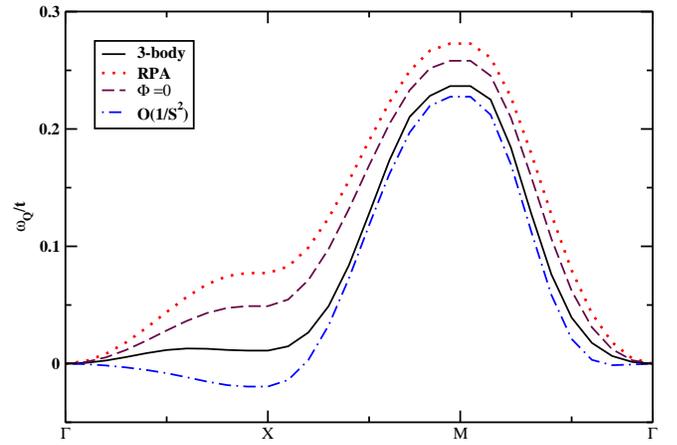}}
}
\caption{
Spin wave dispersion 
along the different directions
in Brillouin zone for $n=0.7, J=8t$: 
Comparison of the different approximations.
}  
\end{figure}

It is important to note here that the strong magnon softening and instability 
only occur for finite values of $J/t$ 
and disappear in the strong coupling limit 
$J \rightarrow \infty$.
This can be seen in Fig. 8, which compares the 
2D magnon dispersions for $n=0.7$ and 
different values of $J/t > 1$ with 
the result obtained by expanding Eqs.(\ref{Psi}) and (\ref{mag-en}) 
in the limit $J \rightarrow \infty$. 
As can be seen in Fig.8, the zone boundary 
magnon softening disappears with increasing 
$J/t$. The magnon dispersions converge slowly to 
the strong coupling result, 
which is reached only for $J/t \sim 1000$. 
Since the typical exchange interaction  values 
in the manganites  
are of the order of $J/t \le 10$, 
we conclude that the manganites are far from the 
$J \rightarrow \infty$ limit.
Noting in Fig.8 that the 
zone boundary magnon softening has 
disappeared completely  for $J/t \ge 20$, 
we see that the finite $J/t$ effects 
play an important role in the manganite spin  dynamics. 

We now turn to the 3D system, 
where the full three--body variational calculation 
faces computational difficulties due to the 
large number of variational parameters 
$\Psi^{{\bf Q}}_{{\bf \alpha} {\bf \nu}}$. 
As can be seen in Fig. 7,  in the 
2D system the 
magnon dispersion results for $n=0.7$ 
have already converged reasonably 
well for $N \sim 10$.
We therefore expect that, in 3D, a  calculation 
for a $N \times N \times N$ lattice 
with $N \sim 10$ 
should give reasonable results.
Fig.9 show the 3D magnon dispersions 
obtained this way for $N=8$, $n\sim 0.7$, and $J/t=14$
using the different approximations.
Fig. 9 shows
similar 3D magnon behavior  as in the 2D system (Fig. 5): 
magnon softening close to the $X$ point
and significant 
deviations between the different approximations 
along $\Gamma-X$ even for 
this relatively large $J/t=14$. 
Similar to the 2D and 1D systems, 
the $O(1/S^2)$ dispersion and the carrier--localized spin scattering 
($\Phi=0, \Psi \ne 0$) variational 
results bound the full three--body magnon dispersion.

\begin{figure}
\centerline{
\hbox{\psfig{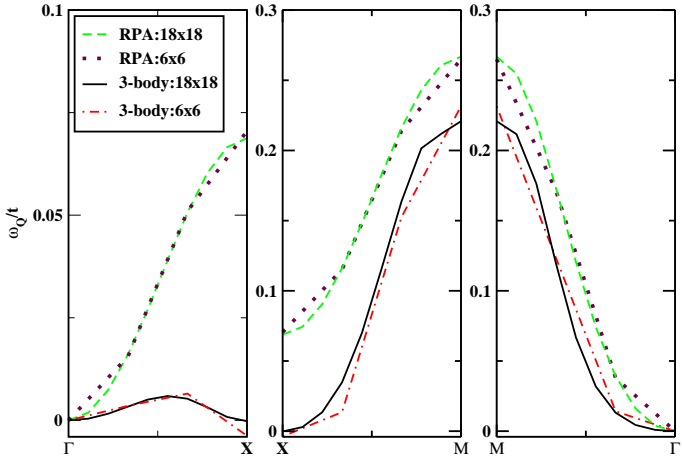}}
}
\caption{
Spin wave dispersion for a $N\times N$  2D lattice  
along different directions
($n=0.7, J=7.5t$).  Convergence with system size $N$
is very good for this $n$.  
}  
\end{figure}

Next we turn to the stability of the fully 
polarized half metallic ferromagnetic state. 
Our results  show 
two different instabilities
due to the exchange interaction ($H_U=H_{super}=0$). 
The first  is the $X$--point 
zone boundary instability discussed above,  
which occurs in 2D and 3D  for intermediate 
electron concentrations.
For low or high electron concentrations
(for all concentrations in 1D),  
we find a second instability  with respect 
to long wavelength spin waves
(negative magnon stiffness) similar to previous calculations. 
In this case,
the minimum magnon energy 
occurs at a finite momentum value instead of 
${\bf Q}=0$. This 
momentum increases with $n$ and becomes 
$\pi$ at  $n=1$ (antiferromagnetic order at 
half filling).
This result implies instability 
to a spiral state, while the 
system can further 
lower its energy 
by phase separating.
\cite{dagotto,papanico} 
Due to the variational nature of our calculation, 
it is guaranteed that, if the magnon energy 
becomes negative for $J = J_c(n)$, 
the ground state of the Hamiltonian $H$ for all  
$J < J_c(n)$  
is not the half metallic state $|F\rangle$.
The phase diagrams of 
Fig. 10 describe  the stability 
of this state against spin wave excitations. 
The most striking feature in Fig. 10 is 
the large shift (increase) 
of the ferromagnetic phase boundary, 
$J_c(n)$, as compared to the RPA, 
due to the  carrier--magnon 
scattering.
Furthermore, 
the different approximations 
of the carrier--magnon 
scattering lead to significant differences in  
$J_c(n)$.
By comparing the shape of 
$J_c(n)$ between the 1D and 2D/3D systems, 
we see that, in the latter case, $J_c(n)$ develops a plateau--like 
shape within an intermediate concentration regime (see Fig. 10(d)). 
This feature is 
absent in the 1D system, 
where there is no zone boundary instability. 
This plateau  
occurs for $0.4 < n \le 0.7$ in 2D
(full three--body calculation) 
and for $0.25 < n < 0.6$ in 3D
($\Phi=0,\Psi \ne 0$ three--body calculation). 
It is much less pronounced for the $O(1/S^2)$ and 
RPA calculations

\begin{figure}
\centerline{
\hbox{\psfig{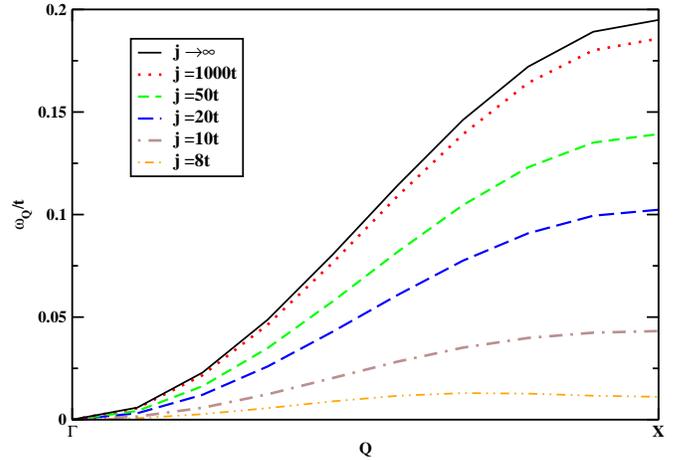}}
}
\caption{
Spin wave dispersion in 2D, obtained from the full three--body calculations,  
for $n =0.7$ and increasing values of $J/t$. 
Convergence to the strong coupling limit is slow. 
}  
\end{figure} 

For small $n$, 
$J_{c}(n)$ is small, 
implying 
enhanced stability of 
the ferromagnetic state in the concentration regime
 relevant, e.g., to III-Mn-V semiconductors.\cite{III-V-1,III-V-2} 
This  stability is a remnant of the fact that,  
in the exactly solvabe limit $N_e=1$, 
the ferromagnetic state $|F\rangle$ is the ground state 
for all values of $J/t$. 
$J_{c}(n)$ increases 
more slowly with $n$ 
in 1D than in  2D and 3D.
This implies 
enhanced stability of the fully polarized ferromagnetic state, 
partly due to the lack of zone boundary instability in 1D.   
Fig. 10(d) shows the 2D phase diagrams   
for the electron concentrations relevant to the manganites.  
The full three--body variational calculation 
gives $J_c(n) \sim 7-8t $ in this regime, 
close to the high end of the
values quoted in the literature. \cite{dagotto} 
Therefore,  the simple 
double exchange Hamiltonian predicts that 
the manganites lie in a regime that is close to the 
instability of the ferromagnetic state. 
In this regime, the  
correlations, vertex corrections, 
and finite $J$ effects play an important role in the spin dynamics.

Fig. 10 also compares the phase diagrams 
obtained by using the different approximations implemented here. 
The $O(1/S^2)$ 
calculation underestimates the stability of the 
fully polarized ferromagnetic state,
while the RPA and carrier--localized spin scattering 
($\Phi=0, \Psi \ne 0$) 
variational results overstimate the stability. 
For low concentrations $n \le 0.3$,  
all the different treatments of the carrier--magnon 
scattering 
predict a similar ferromagnetic phase bounday. 
It is clear from Fig. 10(d) that, in the electron concentration 
regime $0.5 \le n \le 0.8$ 
relevant to the manganites, the RPA significantly overestimates 
the stability of the ferromagnetic state. For example, for 
$n \sim 0.5$,
the RPA underestimates $J_c(n)$ by 100$\%$ as compared to the 
full three--body variational calculation.
Finally, close to half filling $n=1$, 
the two variational results 
give magnon energies similar to the RPA, which  
becomes exact 
for $n=1$. 
On the other hand, the $O(1/S^2)$ approximation fails in this 
high concentration regime.
\begin{figure}
\centerline{
\hbox{\psfig{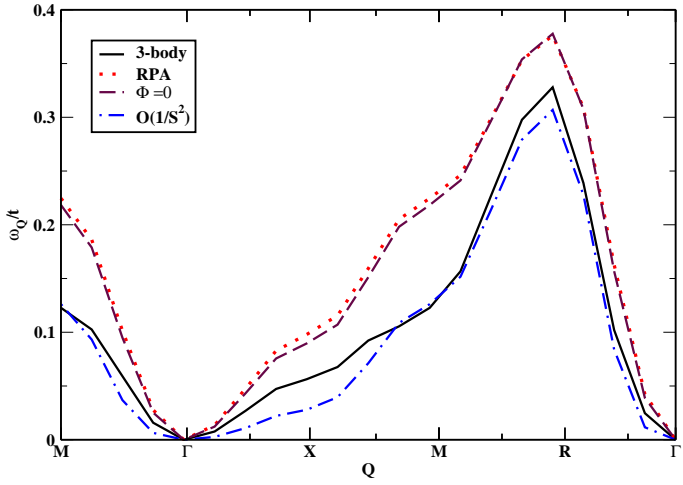}}
}
\caption{
Spin wave dispersion for a $N \times N \times N$  3D lattice  
along different directions
($n=0.7, J=14t$, and $N=8$).
$\Gamma=(0,0,0), X=(\pi,0,0), M=(\pi,\pi,0), R=(\pi,\pi,\pi)$. 
} 
\end{figure}

Finally we discuss the relevance of 
our calculation to the spin wave dispersion observed experimentally 
in the  quasi--2D and 3D manganites.
The experimental results 
are typically analyzed by 
fitting the short range Heisenberg dispersion
to the long wavelength experimental dispersion 
and then comparing 
the two close to the zone boundary.
\cite{sw-exp-1,soft-exp-1,soft-exp-2,soft-exp-3,soft-exp-4,endoh,2D-mang,soft-2D-1,soft-2D-2}  
This comparison  showed that the Heisenberg 
model fails to describe the experimental results 
in the overdoped manganites (typically for $0.5 \le n \le  0.7$), 
but fits well in the underdoped samples 
($n > 0.7$). 
This failure is due to 
the strong magnon softening close to the zone boundary ($X$--point),  
\cite{soft-exp-1,soft-exp-2,soft-exp-3,soft-exp-4,endoh,soft-2D-1,soft-2D-2}  
whose physical origin 
is currently under debate.
\cite{soft-exp-1,soft-exp-2,soft-exp-3,soft-exp-4,endoh,soft-2D-1,soft-2D-2,khal,solovyev,mochena}  
Here we compare
our numerical results with the Heisenberg 
dispersion
$\omega^{Heis}_{{\bf Q}}$,  
obtained  by fitting  
to the long wavelength numerical results, 
by introducing the parameter 
$\Delta =\omega^{Heis}_{X}/\omega_{X}-1,$
where $\omega^{Heis}_{X}$ 
and $\omega_{X}$ are the Heisenberg and numerical magnon energies
calculated at the 
$X$--point.
 $|\Delta|$ thus measures the magnitude of the deviations
from Heisenberg behavior at the zone boundary.  
For example, $|\Delta| \sim 1$ 
means $100 \%$ deviation, 
$\Delta >0$ means magnon softening at the zone boundary,  
as compared to the Heisenberg dispersion with the same stiffness, 
while $\Delta <0$ implies zone boundary hardening. 
Fig. 11 compares  $\Delta(n)$  
obtained from our different approximations. 
With the exception of  small values of $J/t$, 
the RPA gives small deviations 
from Heisenberg behavior, mostly a  hardening at the zone boundary (
$\Delta <0$, see Fig. 10),
and predicts a weak concentration dependence
of $\Delta(n)$. 
This similarity 
between the RPA and Heisenberg dispersions 
is expected for $J \gg t$  since 
the two coincide in the strong coupling limit 
$J \rightarrow \infty$  (see 
Eq.(\ref{RPA-sc})).

\begin{figure}
\centerline{
\hbox{\psfig{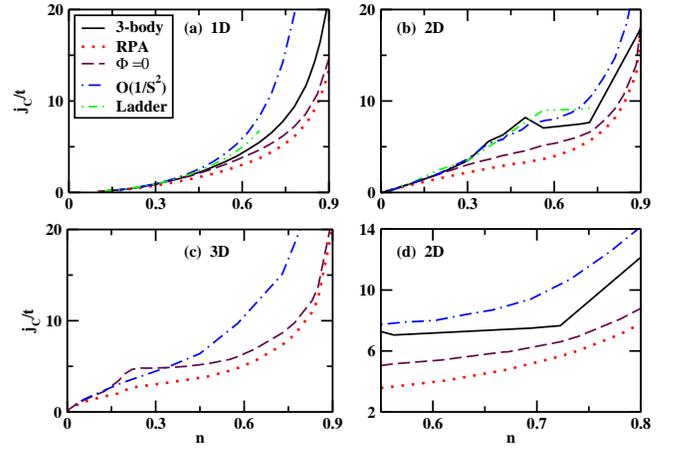}} 
}
\caption{
Phase diagram due to the spin wave instability
and comparison between the different approximations discussed in the text. 
(a) 1D system, (b) 2D system, (c) 3D system, and (d) 2D system in the 
electron concentration range relevant to the manganites.   
}
\end{figure}

The magnon--electron 
scattering leads to larger deviations from Heisenberg ferromagnet spin 
dynamics and enhances $\Delta(n)$  
(see Figs. 11(a) and 11(b) for 2D and 3D respectively). 
In order to compare with the experiment, the value of $J/t$
must be chosen so that $| F \rangle$ is stable 
up to $n \sim 0.8$, 
where a metallic ferromagnetic state is observed experimentally. 
 For $J/t \sim 10$, 
this is the case for the full three--body calculation, 
while larger values of $J/t$ are required to achieve stability 
for $n \sim 0.8$ 
with respect to the $O(1/S^2)$ magnons. 
Figs. 11(a) and 11(b) compare the behavior of $\Delta(n)$
for the different approximations 
in the 2D and 3D systems respectively. 
The $O(1/S^2)$ calculation  
gives magnon {\em hardening} rather than softening in the 
concentration range of interest, similar to the 
strong coupling results of 
Ref. \onlinecite{chubukov}.
This is in contrast to $\Delta(n)$ 
obtained by using  the full  three--body calculation, 
shown in Fig. 11(a) for the 2D system. 
In this case, 
the magnon hardening for $n < 0.5$ 
($\Delta< 0$) 
changes to magnon softening for $0.5 < n < 0.7$ ($\Delta > 0$) 
and then back to a small  magnon hardening for $n > 0.7$. 
This behavior with $n$ is consistent with the 
experimental trends.
Although magnon softening at the $X$ point  can be obtained using 
other  approximations, the full three--body calculation 
gives such an enhanced  effect 
within the range of intermediate 
electron concentrations of interest and for values of  $J/t$ 
such that the fully polarized ferromagnetic state is stable 
for $0.5 \le n \le  0.8$ (where it is 
observed experimentally). 
The above behavior of $\Delta(n)$ is not reproduced 
in the strong coupling limit $J \rightarrow \infty$, 
where magnon hardening is obtained.
It arises from the interplay of the 
$X$--point instability and the 
plateau--like shape of $J_c(n)$, Fig.10, 
induced by the correlations. 
On the other hand, for $J=10t$, 
the carrier--localized spin 
scattering approximation
($\Phi=0, \Psi \ne 0$ 
variational wavefunction)
gives $\Delta(n)$ 
that, 
more or less, follows the RPA behavior
(see Figs. 11(a) and 11(b)).
As $J/t$ decreases,
magnon softening, 
$\Delta > 0$, 
can also be obtained with this approximation 
over a range of electron concentrations 
in both 2D and 3D (see Figs. 11(c) (2D) and 11(d) (3D)). However,  
for such $J/t$, the ferromagnetic state is 
unstable for $n > 0.6$, i.e. in a regime 
where ferromagnetism is observed experimentally. 
We expect that the precise behavior of $\Delta(n)$ 
in the realistic materials 
will also depend on $H_U$, $H_{super}$, and the bandstructure effects
(to be studied elsewhere). 
Here we point out that at least three--body correlations 
must 
be included for a meaningful comparison 
to the experiment. 

\begin{figure}
\centerline{
\hbox{\psfig{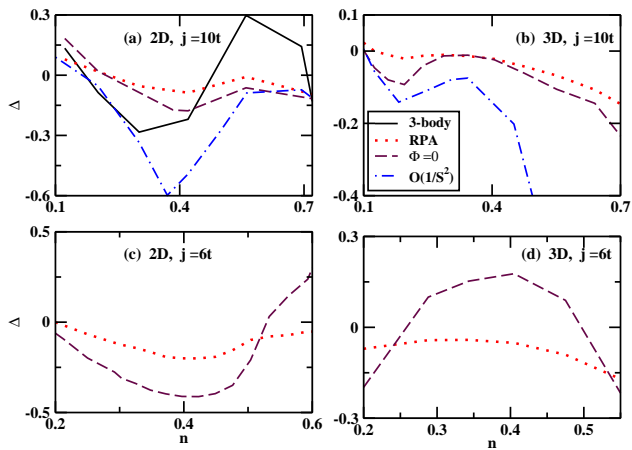}}
}
\caption{Deviation $\Delta(n)$, defined in the text, from 
Heisenberg ferromagnet spin dynamics 
in the 2D and 3D systems for fixed $J/t=10$ 
and different electron concentrations.}
\end{figure}

\section{Conclusions}

In this paper  we presented a nonperturbative variational
calculation of the  
effects of  magnon--Fermi sea pair correlations 
on the spin wave dispersion for the simplest
possible double exchange 
Hamiltonian.
Our theory treats exactly all three--body 
long range correlations between an electron, a Fermi sea hole,  
and a magnon  excitation.
We achieved this 
by using the most 
general 
variational wavefunction that includes up to one Fermi sea pair 
excitations. 
Since the contribution of multipair Fermi sea excitations
is suppressed by powers of $1/S$,
one could alternatively think of our calculation as putting the 
$O(1/S^2)$ result, which treats the one Fermi sea pair 
contribution perturbatively within the Born approximation, 
on a variational nonperturbative basis. 
Our theory  {\bf (i)} becomes exact in 
the two limits  of one and $N^d$ electrons 
and should therefore interpolate well 
 between the low concentration and half filling 
limits, ({\bf ii})  converges well with system size and thus applies to the 
thermodynamic limit, ({\bf iii}) becomes exact in the atomic limit ($t$=0), 
conserves momentum exactly,  
and treats both short and long range correlations on equal basis; 
it should therefore  interpolate well between the 
strong and weak coupling limits, 
 which is important given 
the relatively small values of $J/t$ in the manganites, 
and ({\bf iv}) contains the well known $O(1/S^2)$ and 
RPA results as limiting cases.
In this paper we studied, among others, 
({\bf i}) the shape of the spin wave dispersion 
and ferromagnetic phase boundary 
for different system dimensionalities (1D, 2D, and 3D), 
({\bf ii}) the deviations from the strong coupling double 
exchange limit, and 
({\bf iii}) 
the role of up to three--body 
correlations and nonperturbative vertex corrections 
on the spin dynamics. 
By comparing the full three--body variational calculation 
to a number of approximations (RPA, 1/S expansion, 
ladder diagram treatment of two-body correlations, 
and carrier--localized spin 
rather than  carrier--magnon scattering), we showed that the correlations 
play an important role on the spin excitation 
spectrum, the 
stability of the ferromagnetism, 
 and the shape of the 
ferromagnetic phase boundary  
in the parameter range relevant to the manganites.
Importantly, the correlations lead to spin dynamics
that differs strongly from that of the 
short range Heisenberg ferromagnet 
for intermediate electron concentrations.

Our main results 
can be summarized as follows. 
First, 
the different approximations lead to substantial differences 
in the spin wave dispersion 
and ferromagnetic phase boundary 
for electron concentrations 
above $n \sim 0.3$ and intermediate values 
of $J/t$, which includes the parameter range 
relevant to the 
manganites. 
These large differences come from the correlations, 
which  cannot be neglected, and imply that  
 variational calculations should be used if possible  
in order to draw definite conclusions. 
Second,  we find that,  depending on $n$, 
there are two possibile instabilities 
of the ferromagnetic state toward spin wave excitations: 
instability driven by a negative spin stiffness and
instability at large momenta, close to the $X$--point zone boundary,
with positive stiffness.  
The latter instability only occurs  in the 2D and 3D systems,  
for electron concentrations $n \le 0.7$ and finite values of $J$. 
The 
three--body carrier--magnon correlations
enhance this effect. As a precursor to the above zone boundary 
instabilty, 
we find a strong magnon softening 
at the $X$--point, which should be accompanied by short magnon lifetimes. 
Third,  by comparing to the Heisenberg 
dispersion obtained by fitting 
to the long wavelength numerical results,  
we find strong deviations from the spin dynamics of the short range 
Heisenberg model. 
By choosing  the exchange interaction 
so that the fully polarized ferromagnetic state is stable up to 
$n > 0.8$ as in the experiment, 
we show that the full three--body 2D calculation 
gives strong magnon 
softening 
at the $X$ point for 
$0.5 \le n \le 0.7$, which  
changes into a small hardening for $n > 0.7$. 
This is similar to the  behavior observed in the manganites. 
Our work provides new insight into the spin dynamics in the manganites
and can be extended to treat related ferromagnetic systems (such as e.g. 
the III-Mn-V 
magnetic semiconductors) that are far from the double exchange 
strong coupling limit.
Our calculations imply that the metallic ferromagnetic 
state  in the manganites should be viewed as a 
strongly correlated state.
Finally, the carrier--magnon correlations studied here 
can also play an important 
role 
in the ultrafast relaxation dynamics of itinerant ferromagnetic 
systems, 
which is beginning to be explored by using ultrafast magneto-optical 
pump--probe spectroscopy. 
\cite{ultrafast-per,ultrafast-cmr,ultrafast-ferro} 

We thank N. Papanicolaou for very useful discussions.
This work was supported by the EU Research Training Network  HYTEC
 (HPRN-CT-2002-00315).

\appendix 
\section{}  

\label{append-I}

In this Appendix 
we present the variational equations 
that determine the wavefunction amplitudes 
$X^{{\bf Q}}$, $\Psi^{{\bf Q}}$, 
and $\Phi^{{\bf Q}}$. 
These are obtained by minimizing the 
variational energy 
$\langle F | M_{{\bf Q}} H M^\dag_{{\bf Q}} | F \rangle$, 
where $M^{\dag}_{{\bf Q}}$ is given by Eq.(\ref{magnon}), 
with respect to the above variational amplitudes. 
The normalization condition 
$\langle F | M_{{\bf Q}} M^\dag_{{\bf Q}} | F \rangle =1$ 
is enforced 
via a Lagrange multiplier, which coincides with the 
variational magnon energy $\omega_{{\bf Q}}$. 
Similar to the three--body Fadeev equations, 
the resulting variational equations are equivalent 
to solving the Schr\"{o}dinger equation within 
the subspace spanned by the states 
$S_{{\bf Q}}^{-} | F \rangle$ and 
$c_{{\bf Q}+ {\bf  \nu} \downarrow}^{\dag}
c_{{\bf \nu} \uparrow} | F \rangle$, which 
describe all possible configurations with 
one reversed spin and no Fermi sea excitations, 
and  
the magnon--Fermi sea pair states 
$ c_{{\bf \alpha} \uparrow}^{\dag}
     c_{{\bf \nu} \uparrow} 
S_{{\bf  Q} + {\bf \nu} - {\bf \alpha}}^{-}
| F \rangle$ 
and 
$ c_{{\bf \alpha} \uparrow}^{\dag}
     c_{{\bf \nu} \uparrow} 
      c_{{\bf Q}+ {\bf \mu} + {\bf \nu} - {\bf \alpha} \downarrow}^{\dag} 
      c_{{\bf \mu} \uparrow} | F \rangle$ 
to which a magnon 
or spin flip excitation can scatter with the simultaneous excitation of
a Fermi sea pair. We note that the above  momentum space basis 
ensures the conservation of momentum and total spin.
The explicit form of the variational equations is obtained 
after straighforward algebra by projecting the Schr\"{o}dinger equation 
$[H, M^\dag_{{\bf Q}}] | F \rangle 
= \omega_{{\bf Q}} | F \rangle$ 
in the above basis 
after calculating the commutator 
$[H, M^\dag_{{\bf Q}}]$
by using Eq.(\ref{magnon}) 
for $M^\dag_{{\bf Q}}$
and noting that $H |F \rangle=0$ 
( we take the energy of $|F\rangle$  as zero). 
The variational equation that gives the 
energy $\omega_{{\bf Q}}$  reads 
\begin{eqnarray} 
\omega_{{\bf Q}} = \frac{J n}{2} 
- \frac{J}{2N} \sum_{{\bf \nu}} X^{{\bf Q}}_{{\bf \nu}}
+ \frac{J}{2 N} \sum_{{\bf \alpha} {\bf \nu}} \Psi^{{\bf Q}}_{{\bf \alpha} 
{\bf \nu}}. 
\label{omega-var} 
\end{eqnarray} 
The last term in the above equation describes the contribution 
due to the carrier--magnon scattering. 
The first two terms on the rhs give the RPA magnon energy 
if $X^{{\bf Q}}_{{\bf \nu}}$ is substituted by its RPA
value, 
obtained for $\Psi^{{\bf Q}}=0$. 
The carrier--magnon scattering 
renormalizes $X^{{\bf Q}}_{{\bf \nu}}$
as compared to the RPA result: 
\begin{equation} 
\label{X-var} 
( JS + \varepsilon_{{\bf \nu} + {\bf Q}} 
- \varepsilon_{{\bf \nu}} 
- \omega_{{\bf Q}} ) \, X^{{\bf Q}}_{{\bf \nu}} 
= JS \left[ 1 + \sum_{\alpha} 
\Psi^{{\bf Q}}_{{\bf \alpha} {\bf \nu}}\right].
\end{equation} 
The first term on the rhs of the above equation 
gives the RPA contribution to 
 $X^{{\bf Q}}_{{\bf \nu}}$, while the second term, as well as the correlation 
contribution to the magnon energy $\omega_{{\bf Q}}$, 
describe the effects of the 
magnon--carrier scattering.
The RPA is obtained from Eqs. (\ref{omega-var}) 
and (\ref{X-var})  after setting 
$\Psi^{{\bf Q}}= \Phi^{{\bf Q}}=0$: 
\begin{eqnarray} 
  \omega_{{\bf Q}}^{RPA}  = \frac{Jn}{2}-\frac{J}{2 N}
	  \sum_{{\bf \nu}} X_{{\bf \nu}}^{{\bf Q} RPA}, 
\label{RPAen} 
\\
   X^{{\bf Q} RPA}_{{\bf \nu }}  = \frac{JS}{
JS + \varepsilon_{{\bf \nu} +{\bf Q}}
- \varepsilon_{{\bf \nu}}
- \omega^{RPA}_{{\bf  Q}}}.
\label{RPAX}
\end{eqnarray}
This full RPA result can also be obtained 
as the 
zeroth order contribution to an expansion in powers of 
1/L, where  $L$ is the number of electron flavors 
and corresponding degenerate electron bands.  \cite{papanico} 

The  scattering amplitude $\Psi^{{\bf Q}}$ 
is determined  by the variational equation 
\begin{eqnarray} 
 \left( \omega_{{\bf Q}} - \frac{J n}{2}  
+\varepsilon_{{\bf \nu}}
- \varepsilon_{{\bf \alpha}} 
\right) 
\Psi^{{\bf Q}}_{{\bf \alpha} {\bf \nu}}
= \frac{J}{2N} \left( 1 - X^{{\bf Q}}_{{\bf \nu}} \right)  
\nonumber \\
+ \frac{J}{2N} \sum_{{\bf \alpha}'} 
\Psi^{{\bf Q}}_{{\bf \alpha}' {\bf \nu}}
 - \frac{J}{2N}\sum_{{\bf \nu}'}  \Psi^{{\bf Q}}_{{\bf \alpha} {\bf \nu}'} 
-  \frac{J}{2N} \sum_{{\bf \nu}'}
\Phi^{{\bf Q}}_{{\bf \alpha} {\bf \nu}'{\bf \nu}}. 
\label{Psi-var} 
\end{eqnarray}
The first term on the rhs of the above equation 
gives the Born scattering approximation contribution to the 
carrier--magnon scattering amplitude, 
which is the only one that contributes to $O(1/S^2)$. 
The next two terms describe the effects of the 
multiple electron--magnon (second term) and hole magnon 
(third term) scattering. 
Finally, the last term comes from the 
electronic contribution to the scattered magnon, 
i.e. from the coherent excitation  of a spin--$\uparrow$
electron to the  spin--$\downarrow$
band.
The amplitude $\Phi^{Q}$  of the latter contribution to Eq.(\ref{magnon}) is 
given by the variational equation 
\begin{eqnarray} 
\label{Phi} 
\left( JS + \varepsilon_{{\bf Q} + {\bf \mu} + {\bf \nu} 
- {\bf \alpha}} - \varepsilon_{{\bf \mu}} 
+ \varepsilon_{{\bf \alpha}} - \varepsilon_{{\bf \nu}} 
- \omega_{{\bf Q}} \right) 
\Phi^{{\bf Q}}_{{\bf \alpha} {\bf \nu} {\bf \mu}}
 =  \nonumber \\
J S \left( \Psi^{{\bf Q}}_{{\bf \alpha} {\bf \nu}}
- \Psi^{{\bf Q}}_{{\bf \alpha} {\bf \mu}} \right).
\end{eqnarray} 
We note that, in the 
strong coupling limit $J \rightarrow \infty$, 
$\Phi^{{\bf Q}}_{{\bf \alpha} {\bf \nu} {\bf \mu}}
\rightarrow \Psi^{{\bf Q}}_{{\bf \alpha} {\bf \nu}}
- \Psi^{{\bf Q}}_{{\bf \alpha} {\bf \mu}}$ and 
the last two terms in Eq. (\ref{magnon}) 
describe the 
scattering of a Fermi sea pair 
with the strong coupling 
RPA magnon created by the total spin lowering operator 
$S_{{\bf Q}}^{-}
+\frac{1}{\sqrt{N}} \sum_{{\bf \nu}} 
       c_{{\bf Q}+ {\bf  \nu} \downarrow}^{\dag}
      c_{{\bf \nu} \uparrow}$.
Our general wavefunction Eq. (\ref{magnon})
does not assume an RPA magnon 
and 
includes  corrections to the strong coupling limit that are important for 
the values of $J/t$ relevant to the manganites. 

A closed equation for the  carrier--magnon 
scattering amplitude $\Psi^{{\bf Q}}$ can be obtained 
by substituting in  Eq. (\ref{Psi-var}) 
the expressions for $\Phi^{{\bf Q}}$ 
and $X^{{\bf Q}}$ obtained from Eqs. (\ref{Phi}) and (\ref{X-var}).  
Defining the excitation energy 
\begin{eqnarray} 
\label{D}
\Delta_{{\bf \alpha} {\bf \nu}}^{{\bf Q}}
= \omega_{{\bf Q}} + \varepsilon_{{\bf \nu}}
- \varepsilon_{{\bf \alpha}} \nonumber \\
 - \frac{J}{2N} \sum_{\nu'} 
\frac{\varepsilon_{{\bf Q} + {\bf \nu'} + {\bf \nu} 
- {\bf \alpha}} - \varepsilon_{{\bf \nu}'} 
+ \varepsilon_{{\bf \alpha}} - \varepsilon_{{\bf \nu}} 
- \omega_{{\bf Q}}}{JS + \varepsilon_{{\bf Q} + {\bf \nu'} + {\bf \nu} 
- {\bf \alpha}} - \varepsilon_{{\bf \nu}'} 
+ \varepsilon_{{\bf \alpha}} - \varepsilon_{{\bf \nu}} 
- \omega_{{\bf Q}}} 
\end{eqnarray} 
we thus obtain the following equation:
\begin{eqnarray} 
\Delta^{{\bf Q}}_{{\bf \alpha} {\bf \nu}}
\Psi^{{\bf Q}}_{{\bf \alpha} {\bf \nu}}
= \frac{J}{2N} \frac{\varepsilon_{{\bf \nu} + {\bf Q}}- 
\varepsilon_{{\bf \nu}}- \omega_{{\bf Q}}}{JS + 
\varepsilon_{{\bf \nu} + {\bf Q}}- 
\varepsilon_{{\bf \nu}}- \omega_{{\bf Q}}}
\left[ 1 + 
 \sum_{{\bf \alpha}'} 
\Psi^{{\bf Q}}_{{\bf \alpha}' {\bf \nu}}
\right] \nonumber \\
- \frac{J}{2N}\sum_{{\bf \nu}'}  \Psi^{{\bf Q}}_{{\bf \alpha} {\bf \nu}'} 
\frac{\varepsilon_{{\bf Q}+{\bf \nu} +{\bf \nu}' - {\bf \alpha}} 
- \varepsilon_{{\bf \nu}'
+ \varepsilon_{{\bf \alpha}} 
- \varepsilon_{{\bf \nu}} }
- \omega_{{\bf Q}}}{JS + 
\varepsilon_{{\bf Q}+{\bf \nu} +{\bf \nu}' - {\bf \alpha}}
- \varepsilon_{{\bf \nu}'} 
+ \varepsilon_{{\bf \alpha}} 
- \varepsilon_{{\bf \nu}} 
- \omega_{{\bf Q}}}.
\label{Psi} 
\end{eqnarray} 
The above equation describes up to three--body correlations 
between the magnon and a Fermi sea pair. 
The first term on the rhs describes  the bare carrier--magnon 
scattering amplitude. 
This is renormalized 
by the multiple scattering of a Fermi sea electron 
(second term on the rhs)
and a Fermi sea hole 
(last term on the rhs).  
These two contributions describe   vertex corrections
of the carrier--magnon interaction.  
Eqs.(\ref{omega-var}) and (\ref{Psi}) were solved iteratively until 
convergence for the spin wave energy was reached.

\section{}  

\label{append-II} 

In this Appendix we 
identify the three--body correlation contribution 
to the carrier--magnon scattering amplitude $\Psi^{{\bf Q}}$, 
Eq. (\ref{Psi}), and distinguish it  from the 
two--body multiple scattering contributions. 

We note 
from  Eq. (\ref{Psi}), 
that $\Psi^{{\bf Q}}$
has the form 
\begin{equation}
\label{Psi-G}  
\Psi^{{\bf Q}}_{{\bf \alpha} {\bf \nu}} \
=
\frac{J}{2 N\Delta^{{\bf Q}}_{{\bf \alpha} {\bf \nu}}} \frac{\varepsilon_{{\bf \nu} + {\bf Q}}- 
\varepsilon_{{\bf \nu}}- \omega_{{\bf Q}}}{JS + 
\varepsilon_{{\bf \nu} + {\bf Q}}- 
\varepsilon_{{\bf \nu}}- \omega_{{\bf Q}}}
+ \frac{\Gamma^{e}_{{\bf \nu}} 
- \Gamma^{h}_{{\bf \alpha} {\bf \nu}}}{\Delta^{{\bf Q}}_{{\bf \alpha} {\bf \nu}}},
\end{equation} 
where we introduced the electron 
vertex function Eq.(\ref{Ge}) 
and the hole 
vertex function
\begin{equation} 
\Gamma^{h}_{{\bf \alpha} {\bf \nu}} = 
\frac{J}{2N}\sum_{{\bf \nu}'}  \Psi_{{\bf \alpha} {\bf \nu}'} 
\frac{\varepsilon_{{\bf Q}+{\bf \nu} +{\bf \nu}' - {\bf \alpha}} 
+ \varepsilon_{{\bf \alpha}} 
- \varepsilon_{{\bf \nu}} - \varepsilon_{{\bf \nu}'}
- \omega_{{\bf Q}}}{JS + 
\varepsilon_{{\bf Q}+{\bf \nu} +{\bf \nu}' - {\bf \alpha}}
- \varepsilon_{{\bf \nu}'} 
+ \varepsilon_{{\bf \alpha}} 
- \varepsilon_{{\bf \nu}} 
- \omega_{{\bf Q}}}.
\label{Gh} 
\end{equation} 
Substituting Eq. (\ref{Psi-G}) into Eq. (\ref{Ge}) 
we obtain after some algebra 
that 
\begin{eqnarray} 
\Gamma^{e}_{{\bf \nu}} &=& 
\left[ 1 - 
\frac{J}{2N} \frac{\varepsilon_{{\bf \nu} + {\bf Q}}- 
\varepsilon_{{\bf \nu}}- \omega_{{\bf Q}}}{JS + 
\varepsilon_{{\bf \nu} + {\bf Q}}- 
\varepsilon_{{\bf \nu}}- \omega_{{\bf Q}}}
 \sum_{{\bf \alpha}'} 
\frac{1}{\Delta^{{\bf Q}}_{{\bf \alpha}' {\bf \nu}}} \right]^{-1}  
\times \nonumber \\ 
& & \left[ \frac{J^2}{4N^2} \left( \frac{\varepsilon_{{\bf \nu} + {\bf Q}}- 
\varepsilon_{{\bf \nu}}- \omega_{{\bf Q}}}{JS + 
\varepsilon_{{\bf \nu} + {\bf Q}}- 
\varepsilon_{{\bf \nu}}- \omega_{{\bf Q}}} \right)^2 
\sum_{{\bf \alpha}'}  
\frac{1}{\Delta^{{\bf Q}}_{{\bf \alpha}' {\bf \nu}}} \right. 
\nonumber \\
& & \left.
- \frac{J}{2N} \frac{\varepsilon_{{\bf \nu} + {\bf Q}}- 
\varepsilon_{{\bf \nu}}- \omega_{{\bf Q}}}{JS + 
\varepsilon_{{\bf \nu} + {\bf Q}}- 
\varepsilon_{{\bf \nu}}- \omega_{{\bf Q}}}
\sum_{{\bf \alpha}'} 
\frac{ \Gamma^{h}_{{\bf \alpha}' {\bf \nu}}}{
\Delta^{{\bf Q}}_{{\bf \alpha}' {\bf \nu}}} \right]. 
\label{Ge-eq}  
\end{eqnarray} 
The
first factor on the rhs of the above equation comes from  the 
electron--magnon 
two--body ladder 
diagrams summed to infinite order. The last term in the second factor 
describes the coupling of the electron--magnon and hole--magnon 
scattering channels. This coupling comes from the three--body correlations. 
An analogous equation 
for $\Gamma^{h}$ can be obtained by 
substituting Eq.(\ref{Psi-G}) into Eq. (\ref{Gh}).  
In the case of the simpler variational wavefunction 
$\Phi=0, \Psi \ne 0$, which describes the carrier--localized spin 
scattering contribution, 
the calculation of $\Gamma^h$ simplifies by noting 
from Eq.(\ref{Psi-var}) and the definition Eq.(\ref{Psi-G})   
that $\Gamma^{h}_{{\bf \alpha}}
= \Gamma^{h}_{{\bf \alpha} {\bf \nu}}$.
The corresponding variational equation can be obtained 
by setting $\Phi^{{\bf Q}}=0$ 
in Eq. (\ref{Psi-var}): 
\begin{eqnarray} 
\Gamma^{h}_{{\bf \alpha}} = 
\left[ 1 
+ \frac{J}{2N}\sum_{{\bf \nu}'} 
\frac{1}{\Delta^{{\bf Q}}_{{\bf \alpha}{\bf \nu}'}} \right]^{-1} \times 
\nonumber \\
\left[ \frac{J^2}{4N^2} \sum_{{\bf \nu}'}  
\frac{\varepsilon_{{\bf \nu}' + {\bf Q}}- 
\varepsilon_{{\bf \nu}'}- \omega_{{\bf Q}}}{JS + 
\varepsilon_{{\bf \nu}' + {\bf Q}}- 
\varepsilon_{{\bf \nu}'}- \omega_{{\bf Q}}}
\frac{1}{\Delta^{{\bf Q}}_{{\bf \alpha} {\bf \nu}'}} 
+ \frac{J}{2N}\sum_{{\bf \nu}'} 
\frac{\Gamma^{e}_{{\bf \nu}'}}{\Delta^{{\bf Q}}_{{\bf \alpha} 
{\bf \nu}'}} \right].   \ \ 
\label{Gh-eq} 
\end{eqnarray} 
The first factor on the rhs 
comes from the hole--magnon ladder 
diagrams, while the coupling to $\Gamma^e$ comes from the
three--body correlations.



\begin{references}



\bibitem{nagaev} 
See e.g. E. L. Nagaev, Phys. Rep. {\bf 346}, 387 (2001)
and references therein.

\bibitem{nagaev-HP}
E. L. Nagaev, Phys. Rev. B {\bf 58}, 827 (1998)

\bibitem{nagaev-RKKY}
E. L. Nagaev, Sov. Physics Solid State, {\bf 11} (1970)

\bibitem{III-V-1} See e.g.
 J. K\"{o}nig, J. Schliemann, T. Jungwirth, and A. H. MacDonald, 
in {\em Electronic Structure and 
Magnetism of Complex Materials}, 
eds.  J. Singh and D. A. Papaconstantopoulos (Springer-Verlag, Berlin, 2002).  


\bibitem{III-V-2} M. Berciu and R. N. Bhatt, 
Phys. Rev. B {\bf 66}, 085207 (2002); 
P. M. Krstajic, F. M. Peeters, V. A. Ivanov, V. Fleurov, 
and K. Kikoin, Phys. Rev. B {\bf 70}, 
195215 (2004). 




\bibitem{tokura} 
See e.g. {\em Colossal Magnetoresistance Oxides}, 
ed. Y. Tokura (Gordon Breach, Singapore, 2000)
and references therein. 
 


\bibitem{dagotto} 
E. Dagotto, T. Hotta, and A. Moreo, 
Phys. Rep. {\bf 344}, 1 (2001). 


\bibitem{de}
C. S. Zener, Phys. Rev. {\bf 82}, 403 (1951); 
P. W. Anderson and H. Hasegawa, Phys. Rev. {\bf 100}, 
675 (1955); P. G. de Gennes, Phys. Rev. {\bf 100}, 564 (1955); 
K. Kubo and N. Ohata, 
J. Phys. Soc. Jpn. {\bf 33}, 21 (1972). 


\bibitem{golosov} 
D. I. Golosov, Phys. Rev. Lett. {\bf 84}, 3974 (2000); 
Phys. Rev. B {\bf 71}, 014428 (2005); and references therein.   

\bibitem{chubukov} 
N. Shannon and A. V. Chubukov,
Phys. Rev. B {\bf 65}, 104418 (2002); 
J. Phys. Cond. Matt. {\bf 14}, L235 (2002). 



\bibitem{furukawa} 
N. Furukawa, J. Phys. Soc. Jpn. {\bf 65}, 1174 (1996). 


\bibitem{RPA} 
X. Wang, Phys. Rev. B {\bf 57}, 7427 (1998). 


\bibitem{num-1} 
J. Zang. H. R\"{o}der, A. R. Bishop, and S. A. Trugman, 
J. Phys. Cond. Matt. {\bf 9}, L157 (1997).

\bibitem{num-2} 
T. A. Kaplan and S. D. Mahanti, 
J. Phys. Cond. Matt. {\bf 9}, L291 (1997). 

\bibitem{num-3}
T. A. Kaplan and S. D. Mahanti,
Phys. Rev. Lett. {\bf 86}, 3634 (2000).

\bibitem{okabe} 
T. Okabe, 
Phys. Rev. B {\bf 57}, 403 (1998).

\bibitem{okabe-prog}
T. Okabe,
Theor. Phys. {\bf 97}, 21 (1997).

\bibitem{wurth} 
P. Wurth and E. M\"{u}ller-Hartmann, 
Eur. Phys. J. B {\bf 5}, 403 (1998). 




\bibitem{roth} 
L. M. Roth, Phys. Rev. Lett. {\bf 20}, 1431 (1968); 
Phys. Rev. {\bf 186}, 428 (1969); 
J. Phys. Chem. Solids {\bf 28}, 
1549 (1967).


\bibitem{rucken} 
See e.g. 
A. E. Ruckenstein and S. Schmitt--Rink, 
Int. J. Mod. Phys. B {\bf 3}, 
1809 (1989) and references therein. 


\bibitem{ander} 
B. S. Shastry, H. R. Krishnamurthy, 
and P. W. Anderson, Phys. Rev. B {\bf 41}, 2375 (1990). 

\bibitem{linden} 
W. von der Linden and D. M. Edwards, J. Phys. : Condens. Matter 
{\bf3}, 4917 (1991). 




\bibitem{sw-exp-1} 
T. G. Perring, G. Aeppli, S. M. Hayden, S. A. Carter, 
J. P. Remeika, and S.-W. Cheong, Phys. Rev. Lett. 
{\bf 77}, 711 (1996). 


\bibitem{soft-exp-1}
H. Y. Hwang, P. Dai, S.-W. Cheong, G. Aeppli, D. A. Tennant, 
and H. A. Mook, Phys. Rev. Lett. {\bf 80}, 1316 (1998). 

\bibitem{soft-exp-2}
P. Dai, H. Y. Hwang, 
J. Zhang, J. A. Fernandez-Baca, S.-W. Cheong, C. Kloc, 
Y. Tomioka, and Y. Tokura, 
Phys. Rev. B {\bf 61}, 9553 (2000). 


\bibitem{soft-exp-3}
L. Vasiliu-Doloc, J. W. Lynn, A. H. Moudden, 
A. M. de Leon-Guevara, and A. Revcolevschi, 
Phys. Rev. B {\bf 58}, 14913 (1998). 


\bibitem{soft-exp-4}
T. Chatterji, L. P. Regnault, 
and W. Schmidt, Phys. Rev. B {\bf 66}, 214408 (2002). 



\bibitem{endoh} 
Y. Endoh, H. Hiraka, Y. Tomioka, Y. Tokura, N. Nagaosa, and T. Fujiwara, 
Phys. Rev. Lett. {\bf 94}, 017206 (2005). 



\bibitem{soft-2D-1} 
T. Chatterji, L. P. Regnault, 
P. Thalmeier, 
R. Suryanarayanan, G. Dhalenne, and A. Revcolevschi, 
Phys. Rev. B {\bf 60}, R6965 (1999). 

\bibitem{soft-2D-2} 
N. Shannon, 
T. Chatterji, F. Ouchni, 
and P. Thalmeier, 
Eur. Phys. J. B {\bf 27}, 287 (2002). 



\bibitem{khal} 
G. Khaliullin and R. Kilian, Phys. Rev. B {\bf 61}, 3494 (2000). 



\bibitem{solovyev} 
I. V. Solovyev and K. Terakura, 
Phys. Rev. Lett. {\bf 82}, 2959 (1999). 


\bibitem{mochena} 
S--S Feng and M. Mochena, 
J. Phys. Cond. Matt. {\bf 17}, 3895 (2005); F. Mancini, 
N. B. Perkins, and N. M. Plakida, 
 Phys. Lett. A {\bf 284}, 286 (2001).  



\bibitem{igarashi} 
J. Igarashi, M. Takahashi, and T. Nagao, 
J. Phys. Soc. Jpn {\bf 68}, 3682 (1999); 
J. Igarashi,  J. Phys. Soc. Jpn {\bf 54}, 260 (1985). 





\bibitem{perakis} 
I. E. Perakis and Y.--C. Chang, Phys. Rev. B 
{\bf 47}, 6573 (1993); {\em ibid} 
{\bf 44}, 5877 (1991); and references therein. 

\bibitem{ssr} 
J. F. Mueller, A. E. Ruckenstein, and S. Schmitt--Rink, 
Phys. Rev. B {\bf 45}, 8902 (1992); 
A. E. Ruckenstein and S. Schmitt--Rink, 
Phys. Rev. B {\bf 35}, 
7551 (1987); and references therein.




\bibitem{magnon-2D}
T. G. Perring, D. T. Adroja, G. Chaboussant, G. Aeppli, T. Kimura, 
and Y. Tokura, Phys. Rev. Lett. {\bf 87}, 217201 (2001). 



\bibitem{2D-mang} 
Y. Moritomo, A. Asamitsu, H. Kuwahara, and Y. Tokura, 
Nature {\bf 380}, 141 (1996); H. Martinho, C. Rettori, D. L. Huber, 
J. F. Mitchell, and S. B. Oseroff,
Phys. Rev. B {\bf 67}, 214428 (2003). 

\bibitem{Tc-stiff} 
See e.g. Y. Endoh and K. Hirota, 
J. Phys. Soc. Jpn {\bf 66}, 2264 (1997).  



\bibitem{papanico} M. Marder, N. Papanicolaou, and G. C. Psaltakis, 
Phys. Rev. B {\bf 41}, 6920 (1990); L. R. Mead and N. Papanicolaou,
Phys. Rev. B {\bf 28}, 1633 (1983). 


\bibitem{ultrafast-per} 
J. Chovan, E. G. Kavousanaki, and I. E. Perakis, 
cond-mat/0508178; Phys. Rev. Lett. (to be published).  

\bibitem{ultrafast-cmr} 
Y. H. Ren, X. H. Zhang, G. L\"{u}pke, M. Schneider, 
M. Onellion, I. E. Perakis, Y. F. Hu, and Q. Li, 
Phys. Rev. B {\bf 64}, 144401 (2001). 

\bibitem{ultrafast-ferro} 
M. Vomir, L. H. F. Andrade, L. Guidoni, 
E. Beaurepaire, and J.-Y. Bigot, 
Phys. Rev. Lett. {\bf 94}, 237601 (2005); 
J.-Y. Bigot, L. Guidoni, 
E. Beaurepaire, and P. N. Saeta, 
Phys. Rev. Lett. {\bf 93}, 077401 (2004). 

\end{references}
\end{document}